\documentclass{article}

\usepackage[english]{babel}

\usepackage[letterpaper,top=2cm,bottom=2cm,left=2.5cm,right=2.5cm,marginparwidth=1.75cm]{geometry}

\usepackage{amsmath}
\usepackage{dsfont}
\usepackage{amssymb}
\usepackage{graphicx}
\usepackage{stmaryrd}
\usepackage[colorlinks=true, allcolors=blue]{hyperref}
\usepackage[square, numbers, comma, sort&compress]{natbib}
\usepackage{authblk}

\usepackage{enumitem}

\title{Frequent asymmetric migrations suppress natural selection in spatially structured populations}
\date{}
\author{Alia Abbara\textsuperscript{1,2,*}, Anne-Florence Bitbol\textsuperscript{1,2,*}}
\affil{\textbf{1} Institute of Bioengineering, School of Life Sciences, École Polytechnique Fédérale de Lausanne (EPFL), CH-1015 Lausanne, Switzerland\\
\textbf{2} SIB Swiss Institute of Bioinformatics, CH-1015 Lausanne, Switzerland\\
* Corresponding authors: \href{mailto:alia.abbara@epfl.ch}{alia.abbara@epfl.ch}, \href{mailto:anne-florence.bitbol@epfl.ch}{anne-florence.bitbol@epfl.ch}}

\begin{document}
\maketitle

\begin{abstract}
    Natural microbial populations often have complex spatial structures. This can impact their evolution, in particular the ability of mutants to take over. While mutant fixation probabilities are known to be unaffected by sufficiently symmetric structures, evolutionary graph theory has shown that some graphs can amplify or suppress natural selection, in a way that depends on microscopic update rules. We propose a model of spatially structured populations on graphs directly inspired by batch culture experiments, alternating within-deme growth on nodes and migration-dilution steps, and yielding successive bottlenecks. This setting bridges models from evolutionary graph theory with Wright-Fisher models. Using a branching process approach, we show that spatial structure with frequent migrations can only yield suppression of natural selection. More precisely, in this regime, circulation graphs, where the total incoming migration flow equals the total outgoing one in each deme, do not impact fixation probability, while all other graphs strictly suppress selection. Suppression becomes stronger as the asymmetry between incoming and outgoing migrations grows. Amplification of natural selection can nevertheless exist in a restricted regime of rare migrations and very small fitness advantages, where we recover the predictions of evolutionary graph theory for the star graph. 
\end{abstract}

\section*{Introduction}

Natural microbial populations often present complex spatial structures, where not all organisms are in equal competition. For instance, populations of pathogens are subdivided between different organs during infections~\cite{VanMarle07,Schnell10} and evolve within each host during epidemics~\cite{Bertels19}, commensal bacteria are spread through the gut~\cite{engel2013gut} where they evolve~\cite{Garud19,Frazao22}, and ecosystems are shaped by local resources~\cite{allan2021stream}. Even well-agitated liquid suspensions deviate from idealized well-mixed populations where all organisms are in equal competition~\cite{herrerias2018stirring}. To incorporate spatial structure into population models, early works considered populations divided into several well-mixed subpopulations or demes, with possible migrations between them~\cite{Wright31,Kimura64}. In particular, Maruyama showed that the fixation probability of a mutant is not impacted by spatial structure, under the assumption that migrations are sufficiently symmetric to preserve the overall mutant fraction~\cite{maruyama70, maruyama74}. Note however that even highly symmetric spatial structures can impact mutant fixation probability if extinctions of demes occur~\cite{Barton93}. 

Evolutionary graph theory allows to model complex spatial structures~\cite{lieberman2005evolutionary}. In this framework, one individual is located on each node of a graph, and replacement probabilities are specified along its edges. The state of the population evolves according to the Moran model~\cite{moran1962statistical} using a specific update rule. For instance, in the Birth-death update rule (also known as biased invasion process~\cite{Antal06,Houchmandzadeh11}), an individual is first selected proportionally to its fitness to divide, and then its offspring replaces one of its neighbors on the graph. In the death-Birth update rule (also known as biased voter model~\cite{Antal06,Houchmandzadeh11}), an individual is first selected uniformly at random to die, and then one of its neighbors on the graph is selected proportionally to fitness to divide, and sends its offspring to the empty node. Although these two rules seem very similar, choosing one or the other strongly impacts the evolutionary outcome~\cite{Hindersin15,tkadlec2020limits}. For example, the star graph amplifies natural selection under the Birth-death update rule, but suppresses it under the death-Birth update rule~\cite{Kaveh15,Hindersin15,Pattni15}. Evolutionary graph theory models have been generalized by placing well-mixed demes on graph nodes, rather than single individuals, also using the Moran model with update rules~\cite{Houchmandzadeh11,Houchmandzadeh13,Constable14,yagoobi2021fixation,Yagoobi23}. In all these models, population sizes are strictly constant, and birth and death events are coupled and occur in a specific order. Besides, migration events are either coupled to birth and death~\cite{Houchmandzadeh11,Houchmandzadeh13,Constable14,yagoobi2021fixation,Yagoobi23} or independent from them but symmetric~\cite{Yagoobi23}.

In natural microbial populations, the number of individuals is generally not strictly constant, even though it may be limited e.g.\ by resource availability. Furthermore, there is no imposed order of individual birth and death events. Thus, to make a link with natural situations and with evolution experiments~\cite{Lenski91,Elena03,Good17,Kryazhimskiy12,Nahum15,France19,Chen20,Kassen}, a more universal theoretical description, whose results do not depend on microscopic update rules, is needed. We made a first step in this direction in \cite{marrec2021}, by considering independent events of birth, death, and migration, in a model where deme sizes could fluctuate around a steady-state value. When exchanges between demes are rare, we showed that the star graph can either amplify or suppress natural selection depending on the asymmetry between incoming and outgoing migrations to and from the center. However, the results of \cite{marrec2021} are limited to the rare migration regime, where each deme can be considered as either fully mutant or fully wild-type upon migration events.

Here, we present a new model of spatially structured populations on graphs, directly inspired by the batch culture setups with serial transfers that are used in many evolution experiments~\cite{Lenski91,Elena03,Good17,Kryazhimskiy12,Nahum15,France19,Chen20,Kassen}, including those with spatially structured populations~\cite{Kryazhimskiy12,Nahum15,France19,Kassen}. Our model is formally close to a structured Wright-Fisher model, and allows us to bridge classical population genetics models~\cite{Wright31,Kimura64,maruyama70, maruyama74,Barton93} with evolutionary graph theory~\cite{lieberman2005evolutionary,Hindersin15,tkadlec2020limits}. We investigate the impact of population structure on mutant fixation probability and fixation time. We consider frequent migrations between demes, which can result in mixed states of the demes. We find that in this regime, the star suppresses natural selection and accelerates evolutionary dynamics, provided there is asymmetry between incoming and outgoing migrations to and from the center. More generally, using a branching process approach, we demonstrate that with frequent migrations, all graphs strictly suppress natural selection compared to a well-mixed population, except circulation graphs, where the total incoming migration flow equals the total outgoing one in each deme. In this regime, circulation graphs have no impact on fixation probability. Stochastic simulations confirm our analytical predictions, and show that suppression of selection becomes stronger as the asymmetry between incoming and outgoing migrations grows. Amplification of natural selection can nevertheless exist in a restricted regime of rare migrations, where we recover the results of \cite{marrec2021} and the predictions of evolutionary graph theory for the star. 

\section*{Results}

\subsection*{Deme-structured populations with serial dilutions}

\begin{figure}[htb!]
\begin{center}
\includegraphics[scale=0.6]{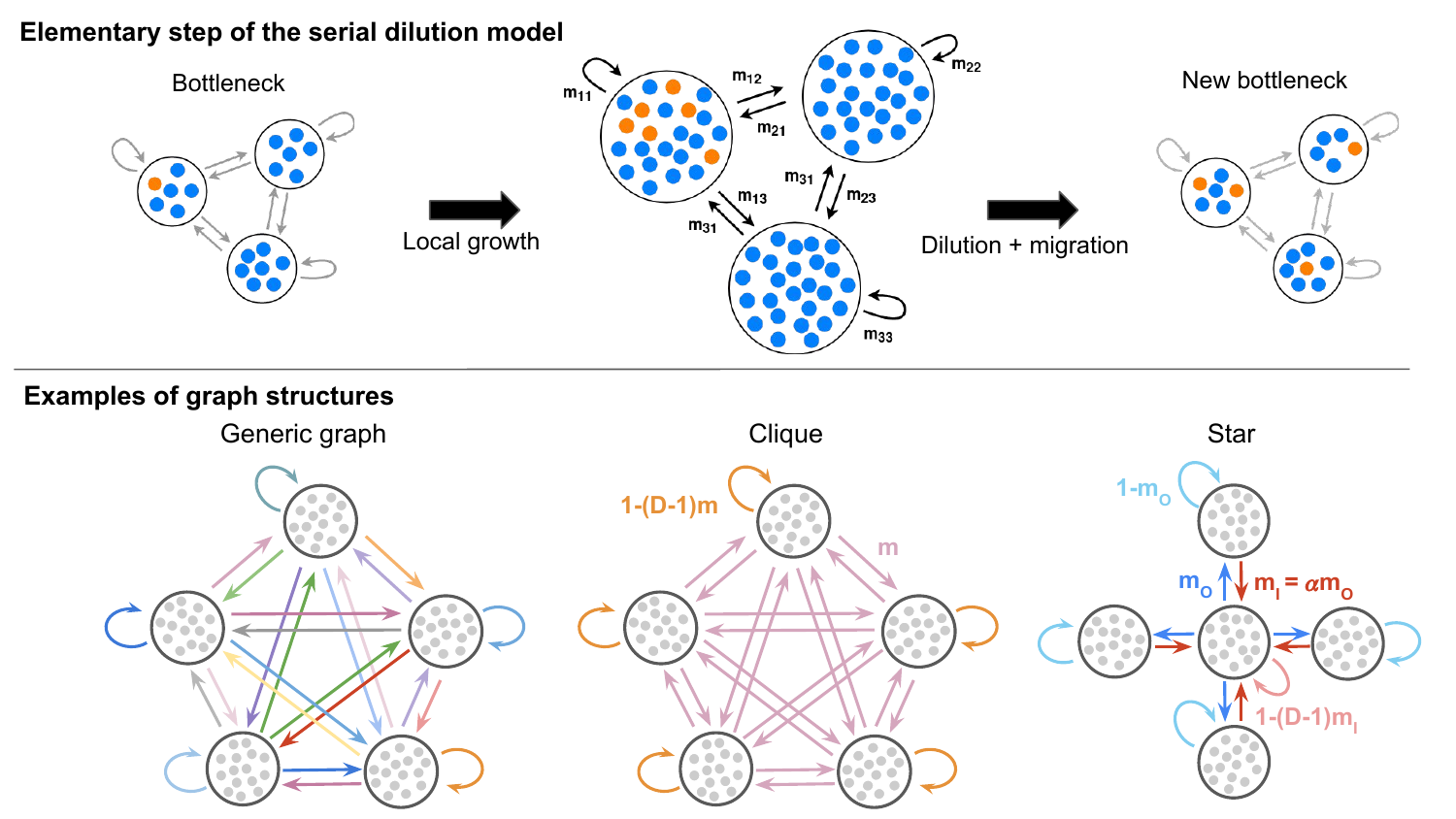}
\caption{\textbf{Schematic of our model and of some graph structures.} Top: one elementary step of the serial dilution model for structured populations. Starting from a bottleneck, demes first undergo a phase of local growth. Then, dilution and migration occur along the edges of the graph, according to migration probabilities. A new bottleneck state is reached.
Bottom: examples of graph structures for $D=5$ demes. From left to right: a generic graph with various migration probabilities, and two strongly symmetric graphs.
For the clique, all migrations probabilities between different demes are equal to $m$. It corresponds to Wright's island model~\cite{Wright31}. For the star, the outgoing migration probability from the center to a leaf is $m_O$, while the incoming migration probability from a leaf to the center is $m_I = \alpha\, m_O$.}
\label{model_structures}
\end{center}
\end{figure}

To model population spatial structure, we consider $D$ demes on the nodes of a connected graph with two types of individuals: wild-types with fitness $f_W=1$, and mutants with fitness $f_M=1+s$. Fitnesses represent division rates during stages of exponential growth. We propose a model with serial phases of exponential growth and dilution (see Methods for details). This model is highly relevant to describe evolution experiments in batch culture with serial transfers~\cite{Lenski91,Elena03,Good17,Kryazhimskiy12,Chen20}, including experiments with controlled spatial structures~\cite{Kryazhimskiy12,Nahum15,France19,Kassen}. In addition, it is formally close to the Wright-Fisher model, allowing us to connect with classical results. For simplicity, individuals are assumed to be haploids that reproduce asexually, but generalizations could be made beyond this case, focusing on a single locus, as in the Wright-Fisher model. The key elementary steps are the following (see Fig.~\ref{model_structures}, top panel). Each deme undergoes exponential growth for time $t$, reaching a large size from an initial bottleneck size. Then, binomial sampling is performed from each deme $i$ to each deme $j$ (including $j=i$) so that on average $Km_{ij}$ individuals are transferred from the grown deme $i$ to form the next bottleneck of deme $j$. Here $m_{ij}$ denotes the probability to migrate from deme $i$ to deme $j$ at a sampling step. Sampling corresponds to dilution and migration. We assume $\sum_i m_{ij}=1$, so the typical bottleneck size of all demes is $K$. The case $\sum_j m_{ij}=1$, where each deme typically contributes by the same amount $K$ to the next bottleneck of the population, is also discussed in the Supplement. This two-step process of growth and dilution-migration is then repeated, until one of the two types of individuals fixes. 

We investigate fixation probability and fixation time in this model. In particular, we ask about the impact of the structure of the graph on these quantities. We consider generic graphs with various migration probabilities, and we give specific results for strongly symmetric graphs, including the clique and the star (see Fig.~\ref{model_structures}, bottom panel). 
 
\subsection*{With frequent migrations, the star suppresses natural selection and accelerates evolutionary dynamics}

The star has been intensely studied in evolutionary graph theory, and is an amplifier of natural selection in the Birth-death process but a suppressor in the death-Birth process~\cite{lieberman2005evolutionary,Antal06, Kaveh15, Hindersin15, Pattni15}. An amplifier of selection yields a higher fixation probability than a well-mixed population for beneficial mutants, and a lower one for deleterious mutants, while a suppressor of selection does the opposite. How does the star impact mutant fixation in our model, which does not rely on an update rule, and where a well-mixed deme sits on each node of the graph? We denote by $\alpha = m_I / m_O$ the asymmetry between incoming probabilities $m_I$ and outgoing migration probabilities $m_O$ between the center and the leaves (see Fig.~\ref{model_structures}). In the restricted regime of rare migrations, we previously showed that migration asymmetry determines whether the star is a suppressor (for $\alpha<1$) or an amplifier (for $\alpha>1$) of selection~\cite{marrec2021}. Here, we consider the more general case of frequent migrations. 

\begin{figure}[htb]
\begin{center}
\includegraphics[scale=0.45]{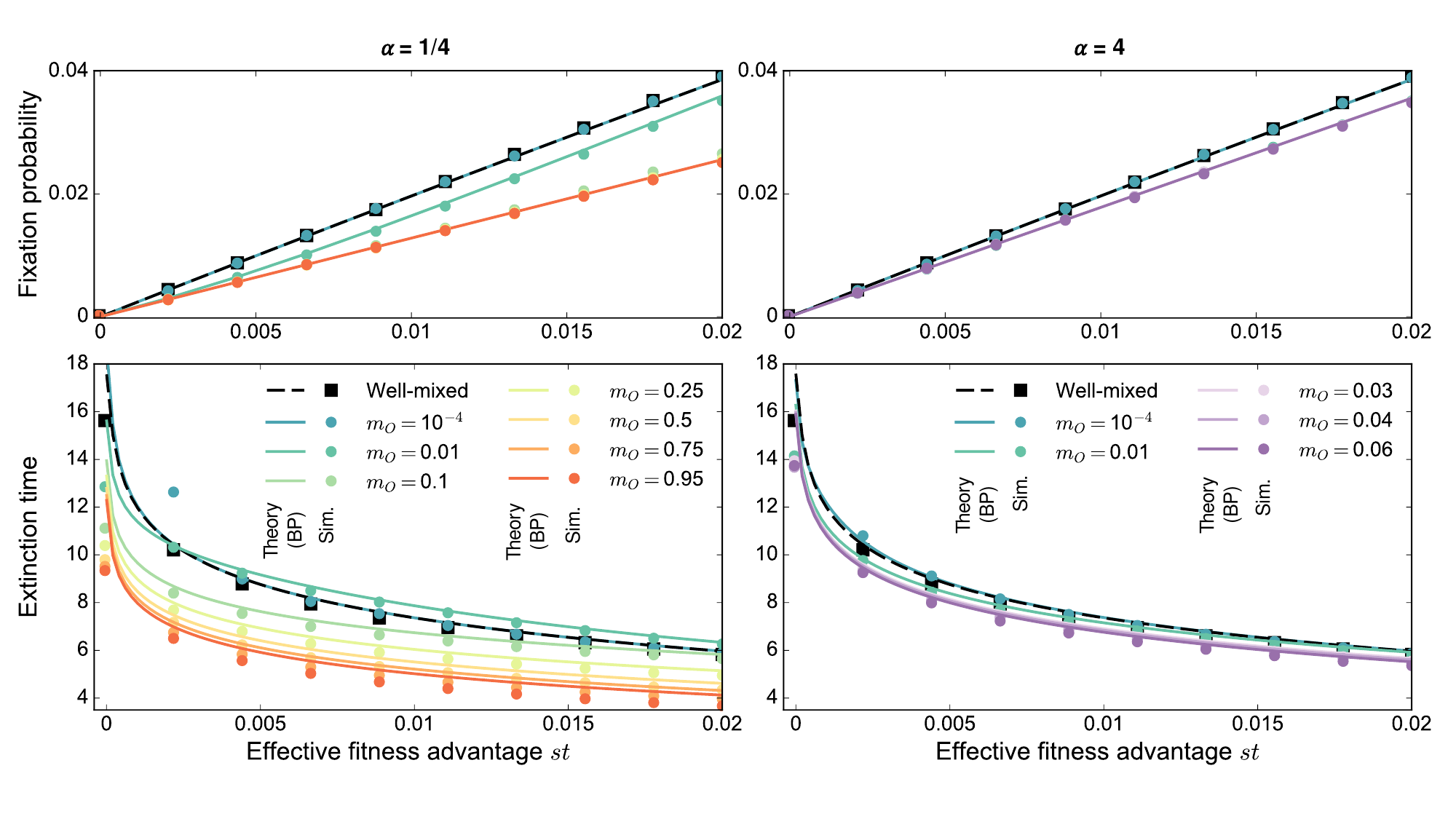}
\caption{\textbf{Mutant fixation and extinction in the star}. Mutant fixation probability (top) and average extinction time (bottom, in numbers of dilution steps) are plotted versus the effective fitness advantage $st$ of the mutant. We consider a star with $D=5$ demes, and $K=1000$ individuals per deme on average at the bottleneck, for migration asymmetries $\alpha=1/4$ (left) and $\alpha=4$ (right).  We start with one mutant of fitness $f_M=1+s$ placed uniformly at random at a bottleneck, all other individuals being wild-types with fitness $f_W=1$. The growth phase duration is $t=5$. Markers represent simulation results (``Sim.''), averaged on 1 million realizations. Lines are theoretical predictions from our branching process (``BP'') approach (see Section~\ref{ex_proba} in the Supplement for fixation probabilities, and Eq.~\ref{tex} for extinction times). The well-mixed case is shown for comparison, with simulations performed for a population with $KD=5000$ individuals at the bottleneck, initialized with one mutant. 
For the star, the outgoing migration probability $m_O$ ranges between $10^{-4}$, above the rare migration regime ($\lesssim 10^{-6}$, see Section~\ref{raremigrations}, esp. Eq.~\ref{rare_condition}, in the Supplement), and values close to 1. Note that for $\alpha=1/4$, $m_O$ can vary between 0 and 1, while for $\alpha=4$, $m_O$ is constrained to a smaller range (see Supplement). }
\label{star_fprobas_etimes_BP}
\end{center}
\end{figure}

Starting with one single mutant placed uniformly at random in a deme at a bottleneck, what is its fixation probability? The coarse-grained description valid for rare migrations, where each deme is either fully mutant or fully wild-type~\cite{Slatkin81,marrec2021}, cannot be used for more frequent migrations. We develop a multi-type branching process approach, which holds when deme size $K$ is large, while the effective fitness advantage $st$ is positive and small, but larger than $1/K$, and for non-rare migrations, see Methods. In what follows, we will refer to this parameter regime as the branching process regime. For the star, we obtain the fixation probability $\rho_C$ (resp. $\rho_L$) starting from a mutant placed in the center (resp. in a leaf) at a bottleneck, as well as their average $\rho=[\rho_C+(D-1)\rho_L]/D$ for a randomly placed mutant, to first order in $st$ (see Supplement, Section~\ref{star_BP_proba}). We find that $\rho_L = \alpha \rho_C$: for $\alpha > 1$, the mutant is more likely to fix starting from a leaf than from the center, and conversely when $\alpha < 1$ (see Supplement, Fig.~\ref{star_fprobas_BP_centerleaf}). More precisely, $\alpha<1$ means that $m_O>m_I$, which makes mutants in the center spread easily to the leaves, giving the center an advantage -- but mutants are more likely to start from the leaves. Moreover, we find that $\rho\leq 2st$ in all cases, where $2st$ is the fixation probability in a well-mixed population~\cite{Haldane27}. Thus, the star always suppresses natural selection in this regime. Fig.~\ref{star_fprobas_etimes_BP} (top panels) shows both analytical predictions and simulation results, in excellent agreement. We observe that while the fixation probability in the star is close to the well-mixed one for relatively rare migrations, suppression becomes stronger as migrations become more frequent.

The average time for mutants to go extinct (conditioned on extinction) can also be derived in our branching process approach (see Methods). We find that extinction becomes faster in the star than in a well-mixed population when migrations are strong. Fig.~\ref{star_fprobas_etimes_BP} (bottom panels) shows excellent agreement between theory and simulations, except for very small values of $st$, which are outside the range of validity of our branching process approach. When migration probabilities are small, exchanges between demes are slow and extinction takes longer than in a well-mixed population. However, as migration probabilities reach $m_O = 0.1$ and above, extinction times become shorter. For such frequent migrations, simulations further reveal a faster growth of mutant fraction in trajectories leading to fixation, and shorter average fixation times, in the star versus the well-mixed population (see Supplement, Figs.~\ref{fraction_evo} and~\ref{star_ftimes}). This is associated to the lower fixation probability in the star: mutants either grow fast enough to survive fluctuations and reach fixation, or go extinct.
These results stand in contrast with those of evolutionary graph theory under the Birth-death process, where spatial structure is generally found to slow down fixation compared to the well-mixed case~\cite{Moller19,Tkadlec19,tkadlec2021fast}. Our model shows such a slowdown of fixation for rarer migrations, in line with expectations (see Supplement, Figs.~\ref{fraction_evo} and~\ref{star_ftimes}).

Qualitatively similar conclusions are obtained if each deme typically contributes by the same amount $K$ to the next bottleneck ($\sum_j m_{ij}=1$ for all $i$), see Fig.~\ref{star_fprobas_etimes_BP_eqcon} in the Supplement. Moreover, different sampling schemes for bottlenecks yield the same results in the branching process regime, see Fig.~\ref{star_fprobas_BP_allsamplings} in the Supplement. This shows the robustness of our conclusions for the star in the branching process regime for strong migrations.


\subsection*{Asymmetry between incoming and outgoing migrations for each deme favors suppression of selection}

Within the branching process approach, and for frequent migrations, we prove that no graph gives a higher fixation probability than the well-mixed population for randomly placed beneficial mutants (see Supplement, Section~\ref{noampli}). In other words, spatial structure cannot amplify natural selection in this regime. Note that simulations show that suppression is still prominent when the fitness advantage of the mutant grows beyond the branching process regime.

The only graphs that do not strictly suppress natural selection for frequent migrations in the branching process regime are such that for each deme, the sum of incoming migration probabilities is equal to the sum of outgoing probabilities (see Supplement, Sections~\ref{circ_thm} and~\ref{strict_supp}). This type of graph is called a \emph{circulation}. Some examples are the clique, and the star with $\alpha=1$, see Fig.~\ref{model_structures}, bottom.
Remarkably, all circulations have the same probability of fixation as well-mixed populations within the branching process approach for frequent migrations (namely, $2st$ to first order in $st$, see Supplement, Section~\ref{circ_thm}). This generalizes the circulation theorem of~\cite{lieberman2005evolutionary} which holds for graphs with one individual per node, as well as our extension~\cite{marrec2021} to graphs with one deme per node in the rare migration regime. Maruyama's pioneering work showed that fixation probabilities are unaffected by spatial structure provided that migration does not change overall mutant frequency~\cite{maruyama70,maruyama74}. This is also known as conservative migration~\cite{Blythe07}. Within our model, since all demes have the same average bottleneck size, i.e.\ $\forall j, \,\,\sum_i m_{ij}=1$, this amounts to requiring that the migration-dilution step preserves the average mutant frequencies in each deme, i.e.\ $x'_i=\sum_k m_{ki}x'_k$, for all possible values of the post-growth mutant fractions $x'_i$. This yields $\forall j, \,\,\sum_i m_{ji}=1=\sum_i m_{ij}$, which corresponds to circulations. Therefore, Maruyama's theorem and the circulation theorem are two faces of a more general result. 

\begin{figure}[htb!]
\begin{center}
\includegraphics[scale=0.7]{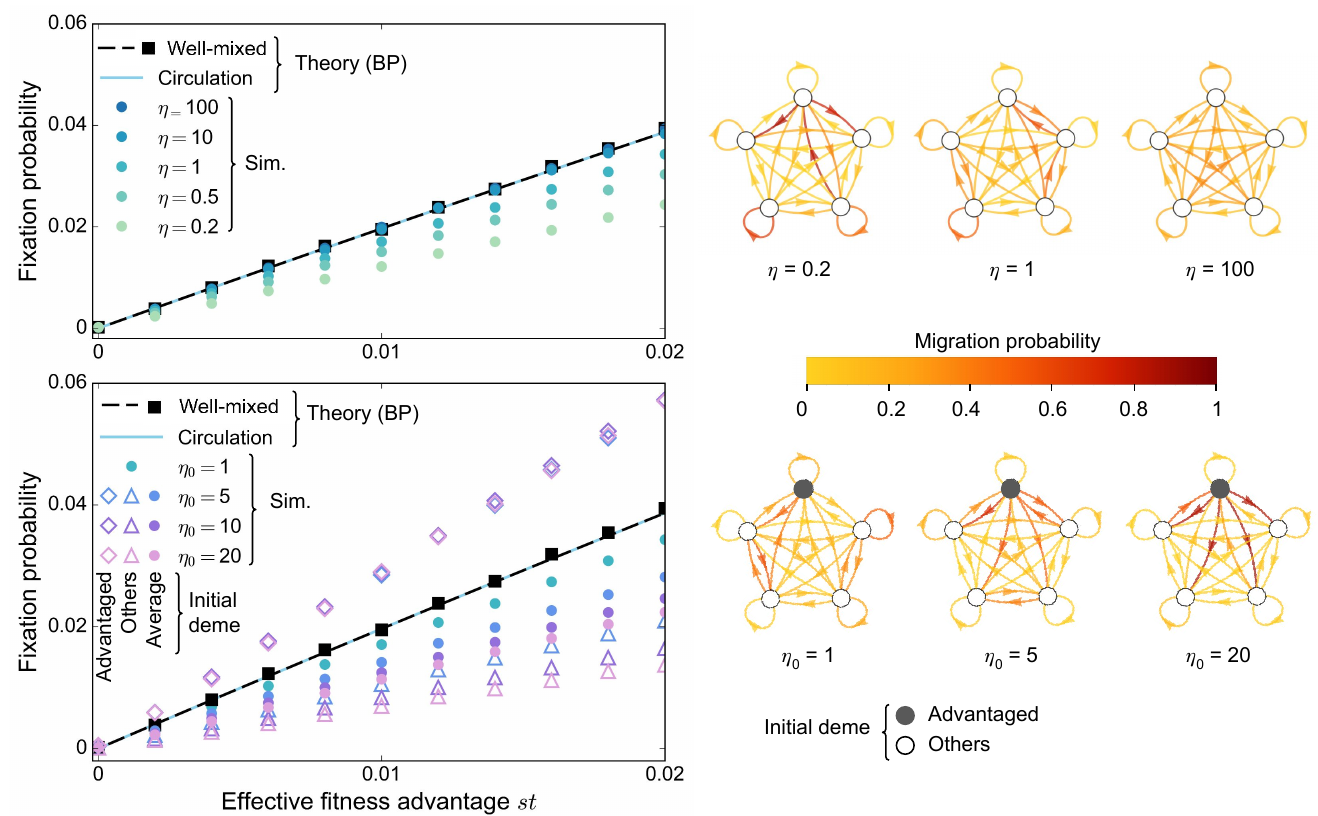}
\caption{\textbf{Mutant fixation probabilities for different asymmetries between incoming and outgoing migrations.} Fixation probabilities are shown versus effective mutant fitness advantage $st$ for Dirichlet cliques (see Supplement, Section~\ref{Dirichlet_clique}) with $D=5$ demes, and $K=1000$ individuals per deme on average at the bottleneck. We start with one mutant of fitness $f_M=1+s$ at a bottleneck, all other individuals being wild-types with fitness $f_W=1$. The growth phase duration is $t=5$. Markers represent simulation results (``Sim.''), averaged on 1 million realizations, with different Dirichlet cliques for each of them. The well-mixed case is shown for comparison, with simulations performed for
a well-mixed population with $KD=5000$ individuals at the bottleneck, initialized with one mutant. Lines are theoretical predictions from our branching process (``BP'') approach, for the well-mixed population and for a circulation with frequent migrations (see Supplement, Section~\ref{ex_proba}). \textbf{Top:} We sample incoming migration probabilities $m_{ij}$ for each destination deme $j$ from the same Dirichlet distribution, using the same parameter $\eta$ for all origin demes. The average value of each $m_{ij}$ is 1/5, but their variances decrease as $\eta$ increases. Examples of generated graphs are shown on the right. Mutants are placed uniformly at random. \textbf{Bottom:} We sample incoming migration probabilities $m_{ij}$ for each destination deme $j$ from a Dirichlet distribution, with parameter $\eta_0 \geq 1$ for one advantaged deme, and 1 for others. Examples are shown on the right. When $\eta_0=1$, all demes are equivalent, recovering the case $\eta=1$ in the top panel. As $\eta_0$ grows, the outgoing migration probabilities from the advantaged deme become larger.
Fixation probabilities are shown when the initial mutant is placed in the advantaged deme, in another deme, and in a deme chosen uniformly at random (``Average'').}
\label{clique_random}
\end{center}
\end{figure}

By contrast, any graph that is not a circulation is a strict suppressor of selection to first order in $st$ for frequent migrations in the branching process regime (see Supplement, Section~\ref{strict_supp}). Indeed, the fixation probability $\rho$ averaged over the initial deme $i$ of the mutant satisfies $\rho <2st$. Graphs deviate from circulations when total incoming and outgoing migrations differ. How does such migration asymmetry impact mutant fixation probability? To investigate this, we consider graphs where this asymmetry can be tuned. Specifically, we generate graphs that we call \emph{Dirichlet cliques}, by sampling all incoming migration probabilities to a given deme $j$ from a Dirichlet distribution, ensuring $\sum_i m_{ij}=1$ (see Supplement, Section~\ref{Dirichlet_clique}). First, we take the same Dirichlet distribution for each destination deme~$j$, with all parameters being equal to $\eta$. All migration probabilities $m_{ij}$ are then centered around the same value, but their variances are tuned by the parameter $\eta$. When $\eta$ is small, migration probabilities have very contrasted values, while they become more homogeneous as $\eta$ grows. A small $\eta$ creates unbalance between the incoming and outgoing migration probabilities for each deme. Fig.~\ref{clique_random} (top panel) shows that Dirichlet cliques generated with large $\eta$ have fixation probabilities very close to those of circulations. As $\eta$ decreases, the probability of fixation in these graphs also decreases on average. Therefore, asymmetry between incoming and outgoing migrations is key to suppression of natural selection, and more asymmetry yields more suppression. 

What happens if one specific deme sends more individuals to other demes than it receives from them? To address this question, we consider Dirichlet cliques where one special deme has a parameter $\eta_0\geq 1$, while all others have $\eta=1$ (see Supplement, Section~\ref{Dirichlet_clique}). Then, $\eta_0$ quantifies the advantage of the special deme: as $\eta_0$ grows, the average value of outgoing migrations from the advantaged deme increases. Consequently, exchanges in the rest of the Dirichlet clique decrease, since $\sum_i m_{ij}=1$ for all $j$. Fig.~\ref{clique_random} (bottom panel) shows that the fixation probability for a mutant starting in the advantaged deme is higher than in a well-mixed population. Indeed, mutants in the advantaged deme can easily spread. Conversely, the spread and thus the fixation of a mutant placed in any other deme is hindered. Averaging over demes, we find that the fixation probability of a mutant placed uniformly at random is smaller than in a well-mixed population. These results generalize those we obtained for the star, where the center is an advantaged deme if $\alpha<1$ (see above and Fig.~\ref{star_fprobas_BP_centerleaf} in the Supplement). Moreover, a stronger unbalance between migration probabilities leads to more suppression of natural selection (see Fig.~\ref{clique_random}, bottom panel).
\newpage
We proved that spatial structure suppresses selection for frequent asymmetric migrations, when mutants are initially placed uniformly at random. In addition, we showed that mutants starting in advantaged demes, which send more individuals to other demes than they receive from them, are more likely to fix than in a well-mixed population, while the opposite holds for other demes. Averaging over all demes can only yield suppression of selection. This is due to the non-linearity (technically to the convexity) of the generating function of the branching process, from which extinction probabilities are derived (see Supplement, Section~\ref{noampli}). Note that the acceleration of mutant extinction in the star versus the well-mixed population (see Fig.~\ref{star_fprobas_etimes_BP}) arises similarly: a slowdown for advantaged demes and an acceleration for others result in an overall acceleration.

\newpage

\subsection*{Amplification can happen for rare migrations and weakly beneficial mutants}

\begin{figure}[htb!]
\begin{center}
\includegraphics[scale=0.48]{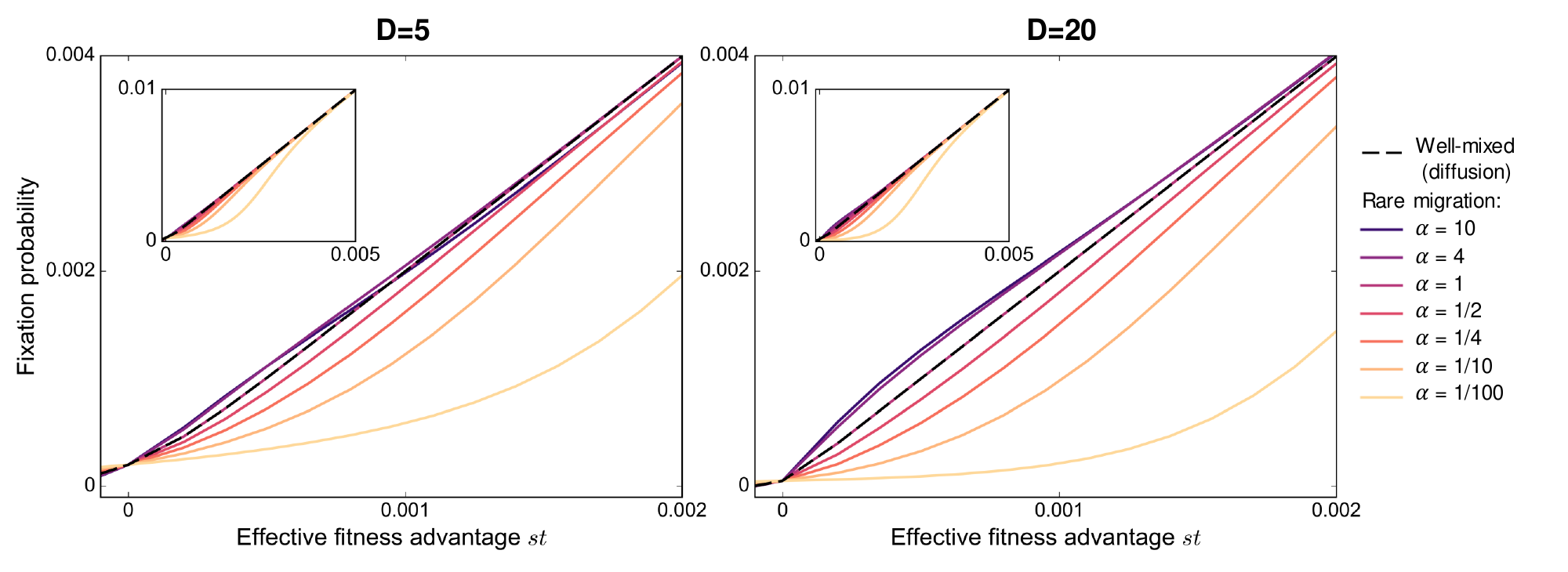}
\caption{\textbf{Mutant fixation probability in the star: rare migration regime.} The analytical fixation probability in the rare migration regime of a single mutant in a star graph is plotted versus effective fitness advantage $st$ with $D=5$ (left) or $20$ (right) demes and $K=1000$ individuals per deme on average at the bottleneck (see Supplement, Section~\ref{fix_proba_star_rare}). Mutants are initially placed uniformly at random at a bottleneck. Results are shown for various values of migration asymmetry $\alpha= m_O/m_I$ (see Fig.~\ref{model_structures}). Black dashed lines give the fixation probability obtained in the diffusion approximation for a well-mixed population with $KD=5000$ individuals initialized with one mutant (see Supplement, Section~\ref{diff_wm}).
Insets show a larger range of effective fitness advantages than main panels.}
\label{rare_migration_star}
\end{center}
\end{figure}

We showed that no graph can amplify natural selection when migrations are frequent, using a branching process approach. How can this be reconciled with the findings of amplification in evolutionary graph theory~\cite{lieberman2005evolutionary}? To address this question, let us consider the rare migration regime in our model. When migrations occur on a longer timescale than the time needed for one mutant to fix in a deme, the graph can be described in a coarse-grained way as having demes that are either fully mutant or fully wild-type~\cite{Slatkin81,marrec2021}. These states can be directly mapped to those of a graph with a single individual per node, as considered in evolutionary graph theory. This mapping breaks down for more frequent migrations, as demes can include various proportions of mutants and wild-types. We studied the rare migration regime in~\cite{marrec2021}, and these results can easily be adapted to our serial dilution model (see Supplement, Section~\ref{raremigrations}). For rare migrations, the star can amplify natural selection if $\alpha>1$, where $\alpha=m_I/m_O$ quantifies the asymmetry between migrations incoming and outgoing to and from the center (see Fig.~\ref{model_structures}). Conversely, it suppresses selection when $\alpha<1$. In particular, starting from a fully mutant deme and for rare migrations, our model exactly maps to evolutionary graph theory under the Birth-death update rule if $\alpha=D-1$~\cite{marrec2021}. Starting from a single mutant, it first needs to fix in its deme before it may spread to other ones. Therefore, its probability of fixation is the product of that in a deme and of that starting from a fully mutant deme. In Fig.~\ref{rare_migration_star}, we show rare migration results for the star starting from a single mutant. We observe that amplification is weak, even though it becomes larger when the number of demes increases. In addition, amplification is restricted to small fitness advantages, see also~\cite{Teimouri23}. Here, we observe that it exists for $st$ of order $1/K$. Increasing the fitness of the mutant makes the fixation probability converge to that of the well-mixed population (Fig.~\ref{rare_migration_star}, insets). 

\begin{figure}[htb!]
\begin{center}
\includegraphics[scale=0.48]{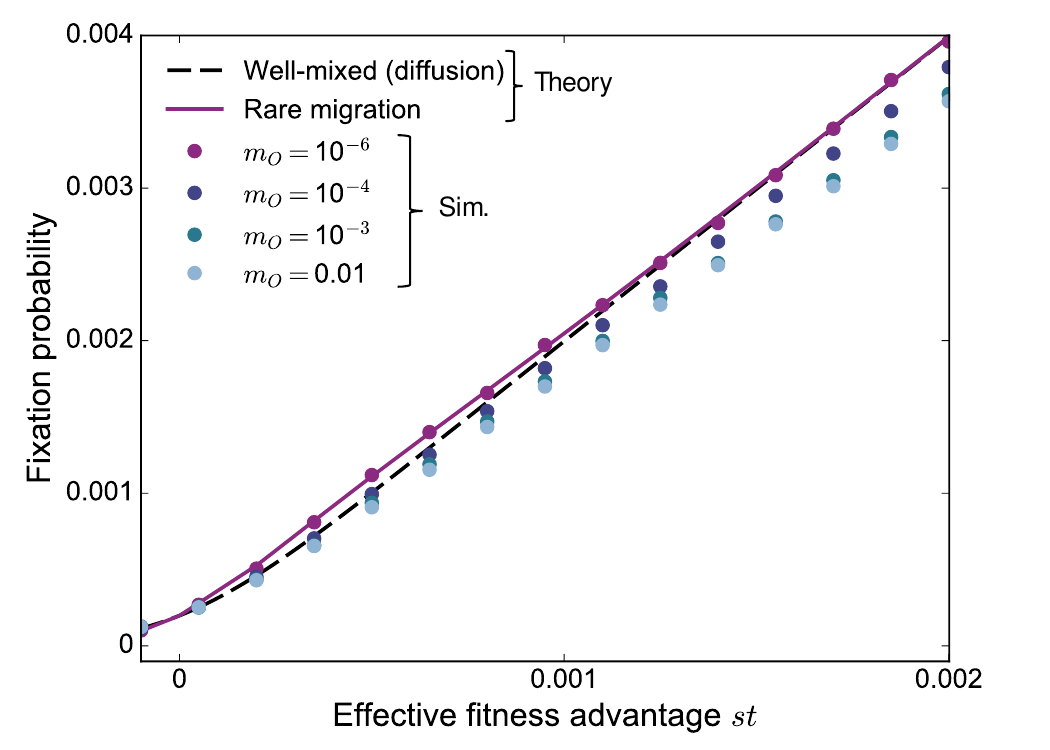}
\caption{\textbf{Mutant fixation probability in the star: from rare to frequent migrations.} The fixation probability of a single mutant in a star graph is shown versus effective fitness advantage $st$ with $D=5$, $K=1000$ and $\alpha=4$. Lines are analytical predictions, shown for rare migrations (see Supplement, Section~\ref{fix_proba_star_rare}), as well as for a well-mixed population of size $KD=5000$ in the diffusion approach  (see Supplement, Section~\ref{diff_wm}). Markers are simulation results (``Sim.'') obtained over 4 million realizations for $m_O=10^{-6}$, and 1 million for other values of $m_O$. For $m_O=10^{-6}$, we are in the rare migration regime, but we exit it as migration probabilities increase above this value (see Supplement, Section~\ref{raremigrations}).}
\label{rare_regime_frequent_migration_star}
\end{center}
\end{figure}

How do our results for frequent migrations connect to those for rare migrations? Simulations allow us to bridge these two regimes. In Fig.~\ref{rare_regime_frequent_migration_star}, we focus on the star with $\alpha=4$, which features amplification of natural selection when migrations are rare. We find that increasing migration probabilities leads to suppression, even in the regime of mutant fitness advantages where amplification exists for rare migrations. In the rare migration regime, a mutant must fix in one deme before it can spread. Since the mutant is placed in a deme chosen uniformly at random, most mutants start in a leaf. When $\alpha>1$, a fully mutant leaf is more likely to send a mutant individual to the center than to receive a wild-type one. This asymmetry provides an extra advantage to a weakly beneficial mutant placed in a leaf. When migrations are frequent, fixation does not occur deme by deme, and this effect disappears.

\newpage

\section*{Discussion}

We have proposed a model of spatially structured populations on graphs, where each node of the graph comprises a well-mixed deme. The population evolves through serial within-deme growth steps and dilution-migration steps. Within a branching process approach, we showed that for frequent migrations, suppression of natural selection is pervasive when there is asymmetry between the total incoming and outgoing migration flows to and from a deme, and becomes stronger when this asymmetry does. Conversely, for circulation graphs where there is no such asymmetry, spatial structure has no effect on mutant fixation probability. 
Our key result that spatial structure suppresses selection for frequent asymmetric migrations holds for mutants initially placed uniformly at random. Mutants starting in advantaged demes, which send more individuals to other demes than they receive from them, are more likely to fix than in a well-mixed population, while the opposite holds for other demes. However, averaging over all demes can only yield suppression of selection, due to the properties of the generating function of the branching process, from which extinction probabilities are derived. In addition to these general results, we studied in detail the star, which can amplify or suppress natural selection depending on the update rule in evolutionary graph theory~\cite{lieberman2005evolutionary,Kaveh15,Hindersin15,Pattni15}. With a deme on each node, the star amplifies selection for rare migrations and small fitness advantages, if incoming migrations to the center are stronger than outgoing ones~\cite{marrec2021}. Here, we showed that it becomes a suppressor of selection for more frequent migrations, provided there is asymmetry between incoming and outgoing migrations to and from the center. In this regime, the star also accelerates evolutionary dynamics compared to a well-mixed population.

Our model establishes a link between classical population genetics models~\cite{Wright31,Kimura64,maruyama70, maruyama74,Barton93} and evolutionary graph theory~\cite{lieberman2005evolutionary,Hindersin15,tkadlec2020limits}. 
Indeed, our finding that circulation graphs do not affect mutant fixation probability is consistent with Maruyama's theorem~\cite{maruyama70, maruyama74} and with the circulation theorem in evolutionary graph theory~\cite{lieberman2005evolutionary}, and connects them. Furthermore, in the rare migration regime, with appropriately chosen migration asymmetries, we recover results from evolutionary graph theory~\cite{marrec2021}. However, we find that amplification of natural selection is limited to rare migrations and small fitness advantages, while suppression of selection is pervasive for frequent migrations, when they are asymmetric.  

The impact of spatial structure on population genetics has often been discussed in terms of effective population sizes. For a given quantity, the effective size of a structured population is the size of a well-mixed population that would yield the same value of this quantity. Thus, the effective size may depend on the quantity of interest~\cite{Ewens79}, and may not exist~\cite{Sjodin05}. Several effective sizes have been employed to characterize spatially structured populations. The inbreeding effective size is associated to the probability of identity-by-descent of two randomly chosen neutral alleles~\cite{Nagylaki80,Whitlock97}. The coalescent effective size may be found by looking for the scaling of time to retrieve the standard coalescent~\cite{Nordborg02,Sjodin05}. In the diffusion approximation, mapping the per-generation mean and variance of the change in mutant frequency to those of a well-mixed population may allow to define an effective selection coefficient and a variance effective population size~\cite{Cherry03}. While these effective population sizes characterize neutral evolution well, they may not suffice to describe evolution under selection~\cite{Barton93, Whitlock97,Nordborg02}. As an illustration, for Wright's island model~\cite{Wright31}, i.e.\ for the clique, the variance effective size is larger than the actual one, but the effective selection coefficient is smaller than the actual one, their product being preserved, consistently with Maruyama's result that fixation probability is unaffected~\cite{Cherry03}. Thus, while at least the inbreeding effective size can be calculated within our model, following~\cite{Whitlock97}, it does not directly shed light on the fixation of beneficial mutants.

Our serial dilution model bridges the gap between theory and experiments. Indeed, batch culture setups with serial transfers are commonly used in evolution experiments~\cite{Lenski91,Elena03,Good17,Kryazhimskiy12,Nahum15,France19,Chen20,Kassen}. Experimentally, it is possible for the dilution step to incorporate exchanges between different demes, thereby allowing to investigate spatially structured populations~\cite{Kryazhimskiy12,Nahum15,France19,Kassen}. Importantly, our results depend on migration asymmetry, which can be directly tuned in such experiments. Experiments with asymmetric migrations were recently initiated in~\cite{Kassen}, in the regime of large fitness advantage of the mutant. We hope that our work will open the way to more quantitative comparisons between theoretical predictions and experimental results for spatially structured populations.

Our branching process approach provides analytical predictions in the regime of large populations sizes, non-rare migrations and small fitness advantage of the mutant. Another important theoretical approach to study populations dynamics, which holds in a larger regime of parameters, is the diffusion approximation~\cite{kimura_diff64}. This approach is well-established for well-mixed populations~\cite{Ewens79}, and has been extended to coupled Wright-Fisher models describing several alleles on multiple loci at linkage equilibrium~\cite{aurell19}. Building upon the link with coalescent theory~\cite{griffiths10,favero21}, these descriptions are the subject of thorough mathematical analysis, and allow for exact simulation methods~\cite{jenkins17, garcia-pareja21}. An interesting perspective would thus be to study our model of spatially structured population within the diffusion approximation, building upon the link with structured Wright-Fisher models~\cite{Lessard07,Burden18}. In particular, it would allow us to study the fate of deleterious mutants. Besides, in this work, we have focused on the fate of mutants that are introduced at a bottleneck. Thus, another extension would be to consider mutants that can appear at any division during the growth phase, building on studies of growth and dilution models for well-mixed populations~\cite{Wahl01,Wahl02,LeClair18,Lin20,Freitas21,Gamblin23}. Beyond the fate of a mutant, investigating how spatial population structure impacts long evolutionary trajectories in our model would be very interesting~\cite{Sharma22}, as well as considering regimes where multiple mutant lineages coexist~\cite{Good17,Blundell19}. Another important extension would be to incorporate changing environments~\cite{Mustonen08,Ashcroft14,Hufton16,Hufton19,Marrec20a,Marrec20b,Marrec23}, and to address cases where demes can go extinct~\cite{Barton93}.
Finally, the impact of spatial structure on mutant fixation is also important in expanding populations. Indeed, the expanding front features reduced effective population sizes and reduced competition. Mutants can then take over by a phenomenon known as gene surfing~\cite{Hallatschek07,Hallatschek08}. Connecting these continuous models of expanding populations to the present discrete models of populations with fixed spatial structure, and addressing population expansion in models on graphs, are interesting topics for future work.

\section*{Methods}

\subsection*{Model of spatially structured populations on graphs with serial dilutions}

In our serial dilution model, we consider a connected graph with $D$ nodes, each comprising a well-mixed deme, and with migration probabilities $m_{ij}$ between each pair of demes $(i,j)\in\{1,\dots,D\}^2$. An elementary step of the dynamics is shown in Fig.~\ref{model_structures} and includes two phases. 

The demes first undergo deterministic exponential growth for time $t$. The growth rates are $f_W=1$ for wild-types and $f_M=1+s$ for mutants. Denote by $M_i$ (resp. $W_i$) the numbers of mutants (resp. wild-types) in deme $i$ at the bottleneck of interest, and by $x_i=M_i/(M_i+W_i)$ the mutant fraction in deme $i$ at this bottleneck. After growth, the total number of individuals is $N'_i=M_ie^t+W_ie^{(1+s)t}$, which is very large (as long as $t$ is not too small), and the fraction of mutants is $x_i'= x_i e^{st} / [1+x_i(e^{st}-1)]$. 

Then, a dilution and migration step is carried out through independent binomial samplings. For each ordered pair of demes $(i,j)$, including $i=j$, two binomial samplings (one for each type, namely mutants and wild-types) take place simultaneously. Each of the $N'_i$ individuals in deme $i$ can be sampled, and each type is sampled proportionally to its frequency after growth. Thus, we sample the number of mutants (resp. wild-types) that migrate from $i$ to $j$ from a binomial law, with $N'_i$ trials, and probability of success $K  m_{ij}x_i'/N'_i$ (resp. $K  m_{ij}(1-x_i')/N'_i$). On average, $K m_{ij}$ individuals migrate from deme $i$ to $j$, resulting into a new bottleneck comprising $K\sum_i m_{ij}$ individuals in deme $j$. Assuming $\sum_i m_{ij}=1$ for all $j$, the average bottleneck size of all demes is $K$. Selection is soft, i.e.\ the contributions of demes are not affected by their average fitnesses~\cite{Wallace75}. Modeling all exchanges between demes through independent binomial samplings allows us to account for fluctuations that would happen at the dilution step in an experiment. 

While the bottleneck size is not strictly fixed in our model, a variant where it is fixed, and where dilution and growth events are performed via multinomial sampling, yields very similar results in the regimes studied here. Note that this multinomial variant can be helpful for small deme sizes, where bottleneck size fluctuations may yield extinctions otherwise. For a single deme, such models with dilution and growth are very close to the Wright-Fisher model, with each bottleneck mapping to a generation~\cite{CrowKimura,manhart2018} (see Supplement, Section~\ref{wmpop}). Therefore, our model is close to a structured Wright-Fisher model. However, note that structured Wright-Fisher models assume binomial sampling within each deme after deterministic migration of offspring~\cite{Lessard07,Burden18}. In Section~\ref{diff_models} of the Supplement, we present two variants of our model, one with multinomial sampling, and the other with binomial sampling within each deme after deterministic migration of offspring. We find the same results with all three variants (see in particular Fig.~\ref{star_fprobas_BP_allsamplings}), which demonstrates the robustness of our conclusions.

We perform stochastic simulations of this model. We also obtain analytical results, using a branching process approach, outlined below.

\subsection*{Branching process analysis}
We describe the state of the population using a multi-type branching process approach, where each type represents each deme, and the number of mutants in each deme is followed~\cite{Harris63,Alexander13}. The branching process description assumes that all mutant lineages are independent~\cite{Harris63}. Under this hypothesis, considering one mutant located in deme $i$ at a given bottleneck, the numbers of its descendants $(n_1,...n_D)$ in the $D$ demes at the next bottleneck follow a probability distribution $\phi_{i}(n_1,...n_D)$. These descendants are the mutants that grew from the initial one in deme $i$, and were then sampled to any destination deme at the migration and dilution step. Assuming independent mutant lineages is valid when mutants are in small numbers, and when deme sizes are all large, i.e. $K\gg 1$. It holds at early phases starting from a single mutant, but fails if the number of mutants becomes large. For mutants with substantial selective advantage, namely for $st\gg 1/K$ here, extinction events happen when mutants are still rare, due to stochastic fluctuations associated to sampling.
Indeed, in a well-mixed population, if their fraction reaches a given threshold, beneficial mutants are very likely to fix in the end~\cite{Desai07,Boenkost21}. Therefore, the branching process approach yields accurate results on extinction probabilities and extinction times provided that $K\gg 1$ and $K st\gg 1$.

Starting from one single mutant in deme $i$, the probability $\phi_i(n_1,...n_D)$ to have $(n_1,...n_D)$ mutants at the next bottleneck is given by the growth and migration-dilution process described above, where mutants migrating to different demes are sampled independently from binomial distributions. We then define a multidimensional generating function $\mathbf{f}$ for $\mathbf{x} \in [0,1]^D$, via its components
\begin{equation}
  f_i(\mathbf{x})=\sum_{n_1,..,n_D=0}^\infty  \phi_i (n_1,...,n_D) \prod_{j=1}^D x_j^{n_j} \quad \text{ for } i = 1,...,D\,.
\end{equation}

Let $p_i$ denote the mutant extinction probability starting from one mutant in deme $i$. The vector of extinction probabilities $\mathbf{p}=(p_1,...,p_D)$ is the only fixed point of $\mathbf{f}$ that is not equal to $(1,...,1)$. To solve the fixed point equation $\mathbf{p}=\mathbf{f}(\mathbf{p})$, we assume $st\ll 1$ (jointly with $K\gg 1$ and $Kst\gg 1$) and write a Taylor expansion of this equation in $st$. This allows to determine the extinction probabilities $p_i$ in this regime, and the fixation probabilities $\rho_i=1-p_i$. 

In the regime of frequent migrations, where all non-zero migration probabilities are much larger than $st$, the binomial distributions used in sampling can be approximated by Poisson ones, see Supplement, Section~\ref{BPPoisson}. For rarer migrations, the full binomial distributions have to be used, and the Taylor expansions need to account for how migration probabilities scale in $st$, see Supplement, Section~\ref{binSampl}. In the Supplement, we explicitly address the case of frequent migrations, as well as the cases where all exchanges between different demes are of order $st$ or $(st)^2$.

Using iterates of the generating function, we can also derive the probability for mutants to be extinct at a given bottleneck, and then the average time to extinction (see Supplement, Section~\ref{ex_time_BP}). 

\section*{Data availability statement} 
All relevant data is included in the main text or in the supplementary material. Python code for our numerical simulations, allowing to reproduce the figures, is freely available at \url{https://github.com/Bitbol-Lab/Structured_pop}.

\section*{Acknowledgments}
 The authors thank Celia García-Pareja and Evan Picchi for helpful discussions. This project has received funding from the European Research Council (ERC) under the European Union’s Horizon 2020 research and innovation programme (grant agreement No.~851173, to A.-F.~B.).

\newpage

\renewcommand{\thefigure}{S\arabic{figure}}
\setcounter{figure}{0}
\renewcommand{\thetable}{S\arabic{table}}
\setcounter{table}{0} 
\renewcommand{\theequation}{S\arabic{equation}}
\setcounter{equation}{0} 

\begin{center}
{\Huge Supplementary Material}
\end{center}
\vspace{1cm}
\tableofcontents

\section{Serial dilution model for a well-mixed population}
\label{wmpop}

Our goal is to develop a serial dilution model for spatially structured populations on graphs. Let us first present the serial dilution model in the simple case of a well-mixed population. We discuss the mutant fixation probability and average extinction time in this model. We do so first within the diffusion approximation (see also \cite{manhart2018}), and next within the coarser branching process approximation.

\subsection{Model}

\paragraph{Hypotheses.} Consider a model with serial dilutions for a well-mixed population of initial bottleneck size $K$. First, it undergoes deterministic exponential growth for a time $t$. Then, $K$ individuals are selected randomly from the grown population to form the next ``generation'', and so on. We assume that there are two types of individuals, wild-types with fitness 1 and mutants with fitness $1+s$. These fitnesses represent deterministic growth rates in the exponential phase. We consider that the sampling is done with replacement from the grown population, and follows a binomial law, as in the Wright-Fisher model.

\paragraph{Variation of the mutant fraction.} Assume that at generation $n$ the initial fraction of mutants (among the $K$ founding individuals of this generation) is $x_n$. After growth, the fraction of mutants reads
\begin{equation}
    x'_n=\frac{x_n e^{st}}{1+x_n(e^{st}-1)}\,.\label{growth}
\end{equation}
Introducing
\begin{equation}
    \sigma= e^{st}-1\,,
    \label{sig}
\end{equation}
we get
\begin{equation}
\label{xpn}
    x'_n=\frac{ (1+\sigma)x_n}{1+\sigma x_n}\,.
\end{equation}

As $K$ individuals are sampled using a binomial law with proportion $x'_n$, the mean shift of the mutant fraction from generation $n$ to generation $n+1$ is
\begin{equation}
\label{Mxn}
    M(x_n)\equiv\langle\delta x_n\rangle=\langle x_{n+1}\rangle-x_n=x'_n-x_n=\frac{\sigma x_n(1-x_n)}{1+\sigma x_n}\approx \sigma x_n(1-x_n)\,,
\end{equation}
where the mean denoted by $\langle .\rangle$ is over the possible values of $\delta x_n=x_{n+1}-x_n$ at a given $x_n$, and where we have assumed $|\sigma|\ll 1$ in the last expression. In addition, the variance of the fraction $x_{n+1}$ of mutant organisms in generation $n+1$ reads
\begin{equation}
    \Delta x_{n+1}^2=\frac{x'_n(1-x'_n)}{K}\,.
\end{equation}
This yields
\begin{equation}
\label{Vxn}
    V(x_n)\equiv\langle(\delta x_n)^2\rangle-\langle\delta x_n\rangle^2=\Delta x_{n+1}^2=\frac{(1+\sigma)x_n(1-x_n)}{K(1+\sigma x_n)^2}\approx\frac{x_n(1-x_n)}{K}\,,
\end{equation}
where we have assumed $|\sigma|\ll 1$ again. 

\subsection{Diffusion approximation} 
\label{diff_wm}
\paragraph{Mutant fixation probability.} Eqs.~\ref{Mxn} and~\ref{Vxn} exactly map to their Wright-Fisher equivalents, with $\sigma$ standing for the fitness advantage of the mutant and $K$ standing for the fixed population size in the Wright-Fisher model. Thus~\cite{CrowKimura}, within the diffusion approximation, which assumes $K\gg1$, the fixation probability $\rho(x)$ of the mutant type starting from a fraction $x$ of mutants satisfies the steady-state Kolmogorov backward equation
\begin{equation}
    \frac{V(x)}{2}\frac{d^2\rho}{dx^2}+M(x)\frac{d\rho}{dx}=0\,, \label{Kol-backward-eq}
\end{equation}
where $M(x)$ and $V(x)$ are given above. In addition, $\rho(0)=0$ and $\rho(1)=1$. Therefore, for $K\gg 1$ and $|\sigma|\ll 1$, we have
\begin{equation}
    \rho(x)=\frac{1-e^{-2K\sigma x}}{1-e^{-2K\sigma}}\,.
    \label{fix0}
\end{equation}
If $K\gg 1$ and $|s|t\ll 1$, combining Eq.~\ref{sig} and Eq.~\ref{fix0} yields
\begin{equation}
    \rho(x)=\frac{1-e^{-2K st x}}{1-e^{-2K st}}\,.
\end{equation}
This exactly matches the Wright-Fisher result, but with $st$ replacing the fitness advantage of the mutant and $K$ standing for the fixed population size in the Wright-Fisher model. In fact, for $t=1$ we recover the Wright-Fisher result with population size $N=K$. 
In particular, starting from one mutant at the bottleneck, $x=1/K$, we have
\begin{equation}
    \rho(1/K)=\frac{1-e^{-2st}}{1-e^{-2Kst}}\,.
    \label{fix}
\end{equation}

\paragraph{Fixation and extinction times.} The average fixation time or extinction time (absorption conditioned on fixation or extinction) starting from an initial frequency $x$ of mutants can also be calculated in the diffusion approximation for the Wright-Fisher process \cite{Kimura1969}. This can be adapted to our serial dilution model.

\subsection{Branching process approximation}

The branching process approximation is coarser than the diffusion approximation, but it will be extremely useful to study our spatially structured populations on graphs. Therefore, here, we present it for the serial dilution model for a well-mixed population.

\paragraph{Hypotheses.} Consider a well-mixed population of bottleneck size $K$ in our serial dilution model, starting from the state where there is 1 mutant and $K-1$ wild-type organisms. In our model, the number of mutants sampled to make generation $n+1$ follows a binomial distribution with parameters $K$ and $x'_n$ (see Eq.~\ref{xpn}). If $K\gg 1$, $x'_n\ll 1$ and $K x'_n$ is of order 1, then this binomial distribution is well approximated by a Poisson distribution with mean $K x'_n$. In particular, at the first generation, with $x_0=1/K\ll 1$, the relevant Poisson distribution has mean
\begin{equation}
    \lambda=K\frac{(1+\sigma)x_0}{1+\sigma x_0}=\frac{1+\sigma}{1+\sigma/K}=1+\sigma-\frac{\sigma}{K}+O\left(\frac{\sigma^2}{K^2}\right).\label{lamWFi}
\end{equation}
Let us assume in addition that $|s|t\ll 1$ and $K|s|t\gg 1$: then, $\sigma=st+(st)^2/2+o((st)^2)$, and $|s|t/K\ll (st)^2$, yielding
\begin{equation}
    \lambda=1+st+\frac{(st)^2}{2}+o\left((st)^2\right)\,.\label{lamWF}
\end{equation}
Assuming that different mutant lineages can be considered independent, which is acceptable if the population is large enough and the mutant fraction remains small enough while the fate of the mutant is set, the number of mutants in the population follows a branching process with offspring distribution given by the Poisson law with mean $\lambda$~\cite{Harris63}. The associated generating function is
$g:x\mapsto\exp\left[\lambda\left(x-1\right)\right]$. 

\paragraph{Mutant fixation probability.} The probability $p$ of extinction of the mutant can be calculated using a branching process approach, as for the Wright-Fisher process~\cite{Haldane27}. Indeed, the extinction probability $p$ is a fixed point of the generating function~\cite{Harris63}, i.e.\ it satisfies 
\begin{equation}
    p=g(p)=\exp\left[\lambda\left(p-1\right)\right]\,.
    \label{genf}
\end{equation}

If $\lambda\leq1$, corresponding to $s\leq0$ (see Eq.~\ref{lamWF}), the only solution of Eq.~\ref{genf} is $p=1$, and extinction is certain. Note that this illustrates that this approach is not taking genetic drift into account properly, and instead focuses on $K\gg 1$. Let us thus focus on the regime where $s>0$, $st\ll 1$, $K\gg 1$ and $Kst\gg 1$. We then expect the extinction probability $p$ to be close to 1. Let us expand it in powers of $st$:
\begin{equation}
    p=1-ast+\frac{b}{2}(st)^2-\frac{c}{6}(st)^3+o((st)^3)\,. \label{pWF}
\end{equation}
To determine $a$ and $b$, let us use Eq.~\ref{genf}, injecting the expansions of $\lambda$ and $p$ from Eqs.~\ref{lamWF} and~\ref{pWF}:
\begin{align}
    1&-ast+\frac{b}{2}(st)^2-\frac{c}{6}(st)^3+o((st)^3)\nonumber\\
    &=\exp\left[\left(1+st+\frac{(st)^2}{2}+o((st)^2)\right)\left(-ast+\frac{b}{2}(st)^2-\frac{c}{6}(st)^3+o((st)^3)\right)\right]\nonumber\\
    &=\exp\left[-ast+\frac{1}{2}\left(b-2a\right)(st)^2+\frac{1}{6}\left(-c+3b-3a\right)(st)^3+o((st)^3)\right]\nonumber\\&=1-ast+\frac{1}{2}\left(b-2a+a^2\right)(st)^2+\frac{1}{6}\left(-c+3b-3a-3ab+6a^2-a^3\right)(st)^3+o((st)^3)\,.
\end{align}
Identifying terms in this expansion in powers of $s$, we obtain $a=2$ and $b=10/3$. Therefore, the fixation probability of the mutant is
\begin{equation}
    \rho=1-p=2st-\frac{5}{3}(st)^2+O((st)^3)\label{WFfixBP}\,.
\end{equation}
To first order in $st$, this is consistent with Eq.~\ref{fix} in the regime where $s>0$, $st\ll 1$, $K\gg 1$ and $Kst\gg 1$. Note however that there is a slight discrepancy on the second order term, whose prefactor is $-2$ in the diffusion approximation and $-5/3$ here.

\paragraph{Average mutant extinction time.} The average mutant extinction time can be obtained from the successive iterates of the generating function~\cite{Harris63}. See Eq.~\ref{tex} below for the general case of the structured population, which can be reduced to the simpler case of the well-mixed population.

\section{Rare migration regime for structured populations on graphs}
\label{raremigrations}
\subsection{Definition of the rare migration regime}

When exchanges between different demes are rare enough, each deme behaves like an isolated well-mixed population, and has time to become completely wild-type or completely mutant, before any migration events occurs. This defines the rare migration regime. Let $t^{(fix)}$ be the fixation time of one mutant in a well-mixed population of bottleneck size $K$, corresponding to that of a single deme. If $m_{ij}$ is the migration probability between different demes $i$ and $j$ at each dilution step, the average number of individuals sent from $i$ to $j$ after $T$ dilution steps is $K m_{ij} T$. In the rare migration regime, no individual is typically exchanged between demes before a mutant has time to reach fixation in a deme. This implies the condition
\begin{align}
    &K m_{ij} t^{(fix)}  \ll 1 \nonumber \\
    & m_{ij} \ll \frac{1}{K t^{(fix)}}. \label{rare_condition}
\end{align}

The average fixation time can be calculated in the diffusion approximation for the Wright-Fisher process \cite{Kimura1969}, and adapted to our serial dilution model.
For example, a mutant of fitness advantage $st=0.01$ fixes in a well-mixed population of bottleneck size $K=1000$ in $t^{(fix)} \approx 7\times 10^2$ dilution steps on average. Therefore, a graph of deme bottleneck size $K=1000$ with $st=0.01$ is in the rare migration regime when all migration probabilites between distinct demes satisfy $ m_{ij} \leq 10^{-6}$.

\subsection{Fixation probability for the star graph}
\label{fix_proba_star_rare}
In the rare migration regime, we use a coarse-grained description of the state of the graph via the state of the demes, which are either completely mutant or completely wild-type~\cite{marrec2021}. The dynamics of the graph is described by a Markov chain: a transition between states can happen when one migration event occur.

Following~\cite{marrec2021}, we can compute the fixation probability starting from one fully mutant deme for a star graph with $D$ demes of size $K$.
Let $\Phi_{(x,i)}$ be the fixation probability when the center is in state $x\in\{0,1\}$ where 0 is wild-type and 1 is mutant, and $i$ leaves are mutant. Then for $i=1,...,D-1$ we have
\begin{align}
\Phi_{(0,i)}&=T_{(0,i)\rightarrow(0,i-1)} \Phi_{(0,i-1)}+T_{(0,i)\rightarrow(1,i)} \Phi_{(1,i)}+[1-T_{(0,i)\rightarrow(0,i-1)}-T_{(0,i)\rightarrow(1,i)}
]\Phi_{(0,i)} \,,\nonumber\\
\Phi_{(1,i-1)}&=T_{(1,i-1)\rightarrow(1,i)} \Phi_{(1,i)}+T_{(1,i-1)\rightarrow(0,i-1)} \Phi_{(0,i-1)}+[1-T_{(1,i-1)\rightarrow(1,i)}-T_{(1,i-1)\rightarrow(0,i)}] \Phi_{(1,i-1)} \,, \nonumber
\end{align}
where the $T$s denote transition probabilities from one state to the other upon a given migration event (see ~\cite{marrec2021} for details).
The above equations become in matrix form
\begin{equation}
    \begin{pmatrix}\Phi_{(0,i)} \\ \Phi_{(1,i)} \end{pmatrix} = \begin{pmatrix}\dfrac{\Gamma_0-\Gamma_1}{1+\Gamma_0} & \dfrac{1+\Gamma_1}{1+\Gamma_0} \\ -\Gamma_1 & 1+\Gamma_1 \end{pmatrix} \begin{pmatrix}\Phi_{(0,i-1)} \\ \Phi_{(1,i-1)} \end{pmatrix}\,,
    \label{matrStar}
\end{equation}
with ${\Phi_{(0,0)}=0}$ and ${\Phi_{(1,D-1)}=1}$. Here we have introduced $\Gamma_0=\gamma/ \alpha$ and $\Gamma_1=\gamma \alpha$, with $\gamma=\rho_W/\rho_M$, where $\rho_M$ (respectively $\rho_W$) is the fixation probability of the mutant starting from one mutant (respectively of the wild-type starting from one wild-type individual) in a well-mixed population of bottleneck size $K$. Meanwhile, $\alpha=m_I/m_O$, where $m_O$ is the migration probability from the center to any leaf while $m_I$ is the migration probability from any leaf to the center.

Solving Eq.~\ref{matrStar} yields the fixation probabilities starting from a fully mutant center, or one fully mutant leaf~\cite{marrec2021}:
\begin{align}
    \Phi_{(1,0)} &= \dfrac{1 - \gamma^2}{1 + \alpha \gamma - \gamma (\alpha + \gamma) \left( \frac{\gamma(1+\alpha \gamma)}{\alpha + \gamma}\right)^{D-1}}\,,\\
    \Phi_{(0,1)} &= \dfrac{\alpha}{\alpha + \gamma}(1 + \alpha \gamma)\Phi_{(1,0)} \,.
\end{align}
The weighted average of these quantities yields the fixation probability starting from a fully mutant deme chosen uniformly at random.
To obtain the fixation probabilities starting from a single mutant, one can just multiply these quantities by $\rho_M$.

Here we employ a serial dilution model while \cite{marrec2021} used a model with individual birth (with logistic rate) and death. Formally, this only affects $\gamma$, in two ways. First, in \cite{marrec2021} it also involved the ratio of the equilibrium wild-type to mutant deme sizes (which was in practice close to one), while here this ratio is one, as the bottleneck size is imposed. Second, in \cite{marrec2021} the fixation probabilities in demes were computed within the Moran model, while here they should be computed within the serial dilution model, i.e.\ using Eq.~\ref{fix} in the diffusion approximation. Apart from these minor differences, the rare migration regime is the same here as in \cite{marrec2021}.

\section{Structured populations on graphs with serial dilutions: branching process with Poisson sampling}
\label{BPPoisson}

\subsection{When all demes start at $K$ on average}

\subsubsection{General description}

\paragraph{Model.} Let us now consider a graph with $D$ nodes, and a well-mixed deme on each of these nodes. This spatially structured population undergoes successive steps of separate exponential growth in each demes and serial dilution and migration events, as described in the main text. We assume that the migration probabilities $m_{ij}$ satisfy $\sum_{i=1}^Dm_{ij}=1$ for all $j$, ensuring that all demes start at an average bottleneck size $K$. We ask about the fate of a mutant that is introduced in deme $i$, with $i\in\{1,\dots,D\}$, at a bottleneck, the rest of the population being fully wild-type. The initial mutant fraction in deme $i$ is $x^{(i)}_0=1/K$. After exponential growth, the mutant fraction in deme $i$ is given by (see Eq.~\ref{growth}):
\begin{equation}
    x'^{(i)}_0=\frac{x^{(i)}_0 e^{st}}{1+x^{(i)}_0(e^{st}-1)}=\frac{\lambda}{K}\,,\label{growthi}
\end{equation}
where $\lambda$ is defined in Eq.~\ref{lamWFi}. Let us assume $s>0$, $st\ll 1$, $K\gg 1$ and $Kst\gg 1$ (as above for the branching process approximation): $\lambda$ is then given by Eq.~\ref{lamWF}.

\paragraph{Poisson approximation.} Upon the first dilution-migration step, we sample on average $K m_{ij} x'^{(i)}_0$ mutants that go to deme $j$ using binomial sampling from the grown deme $i$. The relevant binomial law is $\mathcal{B}\left(N'_i,\,Km_{ij}x'^{(i)}_0/N'_i\right)$, where $N'_i$ is the size of deme $i$ after exponential growth. If $N'_i\gg 1$ and $Km_{ij}x'^{(i)}_0/N'_i\ll 1$ while $\lambda_{ij}=Km_{ij}x'^{(i)}_0$ is of order 1, we can approximate this by the Poisson distribution with mean 
\begin{equation}
    \lambda_{ij}=Km_{ij}x'^{(i)}_0=\lambda m_{ij}\,,
\end{equation}
where we used Eq.~\ref{growthi}. Note that if $m_{ij}$ is of order $st$ or smaller, then $\lambda_{ij}$ is not of order 1: in this case, the Poisson approximation is not justified. (Besides, our upcoming identification of terms in expansions in powers of $st$ will not hold in these cases.) Therefore, let us assume that all $m_{ij}$ that are nonzero are of order 1 (and in particular, $m_{ij}\gg st\gg 1/K$). Note that this is not compatible with the rare migration hypothesis which requires $m_{ij}<1/K$. (Besides, the fact that the rare migration hypothesis requires considering fixation in demes is incompatible with the assumption that the fate of the mutant is set while its fraction remains small, which enters the branching process approximation.) Hence, here we consider the regime of intermediate migrations. In Section~\ref{binSampl}, we will deal with rarer migrations, of order $st$ or $(st)^2$.

\paragraph{Multi-type branching process.}
The full process can be formally described by $\mathbf{Z}^{(0)}, \mathbf{Z}^{(1)},...$, where $\mathbf{Z}^{(k)}$ is a vector that gives the state of our graph at bottleneck step $k$, with the $i$-th component $Z^{(k)}_i$ being the number of mutants in deme $i$. $\mathbf{Z}^{(0)}$ is the initial state of the graph: here, $\mathbf{Z}^{(0)}=\mathbf{e}_i$, the vector with only 0s except one 1 at position $i$, since we initialize with a single mutant in deme $i$ at the bottleneck. Assuming large demes and independence of all mutant lineages, we can use the formalism of multi-type branching processes~\cite{Harris63,Alexander13} to describe the state of our population. Each ``type'' in the branching process corresponds to a different deme in our system. This description generalizes the branching process approximation seen above for the well-mixed population in the serial dilution model.  

\paragraph{Generating function.}
The (multivariate) generating function of our multi-type branching process is a function
$\mathbf{f}: [0,1]^D \rightarrow \mathbb{R}^D $ with $D$ components $f_i : [0,1]^D \rightarrow \mathbb{R}$ defined as follows for all $i \in \{1, \dots, D \}$:
\begin{equation}
\label{genfctgal}
    f_i:\mathbf{x}\mapsto\sum_{r_1,..,r_D=0}^\infty \left(\prod_{j=1}^D x_j^{r_j}\right) \phi_i (r_1,...,r_D) \,,
\end{equation}
where $\phi_i (r_1,...,r_D)$ is the probability that starting from one mutant located in deme $i$, the numbers of mutants at the next bottleneck are $r_1,...,r_D$. In the present case, since 
    \begin{equation}
        \phi_i (r_1,...,r_D)= \prod_{j=1}^D \dfrac{\lambda_{ij}^{r_j}}{r_j!} e^{-\lambda_{ij}} \,,
    \end{equation}
this yields
\begin{equation}
    f_i:\mathbf{x}\mapsto\exp \left[ \lambda  \sum_j m_{ij} (x_j-1)\right]. \label{gen_Poisson}
\end{equation}
As for regular branching processes, the generating function of $\mathbf{Z}^{(k)}$ is the k-th iterate of $\mathbf{f}$. In other words, it is $\mathbf{f}^{(k)}$, with components $f^{(k)}_i$ defined as
\begin{equation}
    f^{(k)}_i (\mathbf{x})= f_i (\mathbf{f}^{(k-1)}(\mathbf{x})) \quad \forall i=1,...,D.
\end{equation}

\paragraph{Matrix of first moments.}
Let $M$ be the matrix of first moments, defined by its elements 
\begin{equation}
    M_{ij}=\dfrac{\partial f_i}{\partial x_j} (1,...,1).
\end{equation}
Computing this matrix yields $M_{ij} = \lambda m_{ij}$.
Recall that $\lambda = 1 + st + O((st)^2) >1$ (as we assume $s>0$). In Section~\ref{eigval1}, we will show that the migration matrix has 1 as largest eigenvalue. Therefore the largest eigenvalue of $M$ is $\lambda$. This eigenvalue affects the properties of the branching process~\cite{Harris63}.

\paragraph{Extinction probabilities.}
Let $p_i$ be the probability of extinction of the mutant, starting from one initial mutant in deme $i$, i.e. $\mathbf{Z}^{(0)}=\mathbf{e}_i$. Let $\mathbf{p}=(p_1,...,p_D)$ be the vector composed of extinction probabilities starting from each deme. As the matrix of first moments $M$ has eigenvalue $\lambda >1$, Theorem 7.1 from \cite{Harris63} states that:
\begin{itemize}[noitemsep,topsep=0pt]
\item $\mathbf{p}$ is a fixed point of the generating function: $\mathbf{p} = \mathbf{f}(\mathbf{p})$,
\item $\forall i, \quad 0 \leq p_i < 1$.
\end{itemize}

\paragraph{Probability of extinction by generation $n$.} The probability of having no mutants left \emph{by} generation $n$ is~\cite{Harris63}
\begin{equation}
    P[\mathbf{Z}^{(n)}=\mathbf{0}]=\mathbf{f}^{(n)}(\mathbf{0}). \label{ex_n}
\end{equation}
Besides, the probability of extinction is the limit when $n \rightarrow \infty$ of the iterated generating function:
\begin{equation}
    \forall i, \quad \lim_{n \rightarrow \infty} f^{(n)}_i(\mathbf{0}) = p_i.
\end{equation}
The $i$-th component of $\mathbf{f}^{(n)}(\mathbf{0})$ grows monotonically towards $p_i$.

\paragraph{Average extinction time.} 
\label{ex_time_BP}
The probability of becoming extinct \emph{at} time $n$ is 
\begin{equation}
    P[\mathbf{Z}^{(n)}=\mathbf{0} | \mathbf{Z}^{(n-1)} \neq \mathbf{0}] = f^{(n)}_i(\mathbf{0})-f^{(n-1)}_i(\mathbf{0}).
\end{equation}
The vector of average extinction times (in numbers of generations) $\mathbf{t}^{(ex)}=(t^{(ex)}_1,...,t^{(ex)}_D)$, where $t^{(ex)}_i$ is the average extinction time when we start with one mutant in deme $i$, is given by
\begin{equation}
    \mathbf{t}^{(ex)}= \mathbf{f(\mathbf{0})}+\sum_{n=1}^\infty n [\mathbf{f}^{(n)}(\mathbf{0}) - \mathbf{f}^{(n-1)}(\mathbf{0})].
    \label{tex}
\end{equation}

\subsubsection{Spatial structure cannot amplify selection within the branching process regime}
\label{noampli}

Within the branching process approach with Poisson sampling, let us show that the mutant fixation probability in a spatially structured population is never larger than in the well-mixed population. For this, let us start from Eq.~\ref{ex_n}, which gives the probability of extinction of mutants by bottleneck step $k$, starting from a single mutant.
In a well-mixed population, this probability is $g^{(k)}(0)$, where $g(x)=\exp[\lambda(x-1)]$ is the generating function given by Eq.~\ref{genf}. For a structured population, the probability of extinction by bottleneck step $k$ starting from an initial mutant randomly placed in the graph is $\sum_i f_i^{(k)}(\mathbf{0})/D$, where $f_i^{(k)}$ is the $i$-th component of the $k$-th iterate of the generating function defined in Eq.~\ref{gen_Poisson}. Let us show by induction that for any graph, for all $k \geq 1$, $ \sum_i f_i^{(k)}(\mathbf{0})/D\geq g^{(k)}(0)$.

\paragraph{First bottleneck.}
The probability of being extinct at the first bottleneck step in the well-mixed population is $g(0)=e^{-\lambda}$. For a structured population, it reads
\begin{equation}
    \dfrac{1}{D} \sum_i f_i (\mathbf{0}) = \dfrac{1}{D} \sum_i\exp \left( - \lambda \sum_j m_{ij}\right) \geq \exp \left( - \dfrac{\lambda}{D} \sum_i \sum_j m_{ij}\right) = e^{-\lambda}=g(0)\,,
\end{equation}
where we have used the convexity of the exponential function, and $\sum_i m_{ij} = 1$ for all $j$. Thus, we have shown that $ \sum_i f_i(\mathbf{0})/D \geq g(0)$. Therefore, the extinction probability at the first step is larger in the structured population than in the well-mixed one.

\paragraph{From bottleneck step $k$ to $k+1$.}
Assume $g^{(k)}(0)\leq \sum_i f_i^{(k)}(\mathbf{0})/D$ for some $k \geq 1$. The probability that the mutant is extinct by bottleneck step $k+1$ in the well-mixed population is
\begin{equation}
    g^{(k+1)}(0)=g(g^{(k)}(0))=\exp\left\{\lambda \left[g^{(k)}(0)-1\right]\right\}. 
\end{equation}
For a structured population, it reads
\begin{equation}
    \dfrac{1}{D}\sum_i f_i^{(k+1)}(\mathbf{0}) = \dfrac{1}{D} \sum_i \exp \left\lbrace \lambda \sum_j m_{ij} \left[f_j^{(k)}(\mathbf{0})-1\right] \right\rbrace \,,
\end{equation}
and using the convexity of the exponential function and then the induction hypothesis at  bottleneck step $k$, we obtain:
\begin{align}
   \dfrac{1}{D} \sum_i \exp \left\lbrace \lambda \sum_j m_{ij} \left[f_j^{(k)}(\mathbf{0})-1\right] \right\rbrace &\geq \exp \left\lbrace \dfrac{\lambda}{D} \sum_j \sum_i m_{ij} \left[f_j^{(k)}(\mathbf{0})-1\right]\right\rbrace \\
   &\geq \exp \left\lbrace \dfrac{\lambda}{D} \sum_j \left[f_j^{(k)}(\mathbf{0})-1\right] \right\rbrace\\
   &\geq \exp \left\lbrace \lambda \left[g^{(k)}(0)-1\right] \right\rbrace \,.
\end{align}
The induction hypothesis is thus verified at bottleneck step $k+1$. Therefore, we have shown by induction that for any graph, for all $k \geq 1$, $\sum_i f_i^{(k)}(\mathbf{0})/D\geq g^{(k)}(0)$.

\paragraph{Infinite time limit.}
As $k \rightarrow \infty$, we still have
\begin{equation}
     \lim_{n \rightarrow \infty} \frac{1}{D} \sum_i f_i^{(k)}(\mathbf{0})\geq \lim_{k \rightarrow \infty} g^{(k)}(0) \,.
\end{equation}
Therefore the probability of extinction starting from a single mutant is always larger in a structured population than in a well-mixed population. In other words, there can be no amplification of natural selection in a structured population. This proof holds in the branching process regime and for frequent migrations.

\subsubsection{Expansion of the extinction probability}

Let us focus on the probability $p_i$ that the mutant type gets extinct, starting from one single mutant in deme $i$ at the bottleneck. The results in the previous subsection ensure that for all $i$, $p_i$ satisfies:
\begin{equation}
    p_i=\exp\left[\lambda\sum_{j=1}^D m_{ij} \left(p_j-1\right)\right]\,.
    \label{genfsub}
\end{equation}

Under the assumptions $s>0$, $st\ll 1$, $K\gg 1$ and $Kst\gg 1$, let us expand $p_i$ perturbatively in powers of $st$:
\begin{equation}
    p_i=1-a_ist+\frac{b_i}{2}(st)^2-\frac{c_i}{6}(st)^3+o((st)^3)\,, \label{pieq}
\end{equation}
where $a_i$, $b_i$ and $c_i$ are unknown prefactors. To determine $a_i$ and $b_i$ for each deme $i$, let us use Eq.~\ref{genfsub}, using Eq.~\ref{pieq} and recalling that $\lambda= 1+st+(st)^2/2+o((st)^2)$:
\begin{align}
    p_i&=\exp\left[\left(1+st+\frac{1}{2}(st)^2+o((st)^2)\right)\sum_{j=1}^D m_{ij}\left(-a_jst+\frac{b_j}{2}(st)^2-\frac{c_j}{6}(st)^3+o((st)^3)\right)\right]\nonumber\\
    &=\exp\left[-st\sum_{j=1}^D m_{ij}a_j +(st)^2\sum_{j=1}^D m_{ij}\left( \frac{b_j}{2}-a_j\right)+ (st)^3\sum_{j=1}^D m_{ij}\left(\frac{b_j}{2}-\frac{a_j}{2}-\frac{c_j}{6}\right) +o((st)^3)\right]\nonumber\\
    &=1-st\sum_{j=1}^D m_{ij}a_j +\frac{1}{2}\left[\sum_{j=1}^D m_{ij}b_j-2\sum_{j=1}^D m_{ij}a_j+\left(\sum_{j=1}^D m_{ij}a_j\right)^2\right](st)^2+\left[- \frac{1}{6} \left(\sum_{j=1}^D m_{ij}a_j\right)^3\right.\nonumber \\ &+ \left.\sum_{j=1}^D m_{ij}\left(\frac{b_j}{2}-\frac{a_j}{2}-\frac{c_j}{6}\right)  - \left(\sum_{j=1}^D m_{ij}a_j\right) \left(\sum_{j=1}^D m_{ij}\left(\frac{b_j}{2}-a_j\right)\right) \right](st)^3  +o((st)^3) \,.
    \label{piexp}
\end{align}
Combining Eqs.~\ref{pieq} and~\ref{piexp} and identifying terms in these two expansions in powers of $st$, we obtain:
\begin{align}
    a_i&=\sum_{j=1}^D m_{ij}a_j\,,\label{cd1}\\
    b_i&=\sum_{j=1}^D m_{ij}b_j-2\sum_{j=1}^D m_{ij}a_j+\left(\sum_{j=1}^D m_{ij}a_j\right)^2=\sum_{j=1}^D m_{ij}b_j-2a_i+a_i^2\,,\label{cd2}\\
    c_i&=\sum_{j=1}^D m_{ij}c_j-3 b_i + 3 a_i b_i - 3 a_i + 3 a_i^2 - 2 a_i^3 \label{cd3}\,.
\end{align}

\subsection{When all demes contribute by $K$ on average }

So far, we assumed that the migration probabilities $m_{ij}$ satisfy $\sum_{i=1}^Dm_{ij}=1$ for all $j$, ensuring that all demes start at an average bottleneck size $K$. This is the convention taken in the main text. However it is not the only possible choice. Let us now consider the case where $\sum_{j=1}^Dm_{ij}=1$ for all $i$, ensuring that all demes contribute by $K$ on average.

As before, consider the initial condition where 1 mutant is present at the bottleneck in deme $i$, and the rest of the population is fully wild-type. The difference with the previous case is that the size of deme $i$ at the bottleneck is not $K$ but $N_i=K\sum_{k=1}^D m_{ki}$. Thus, the initial mutant fraction in deme $i$ reads $x^{(i)}_0=1/N_i$, and after exponential growth, the mutant fraction in deme $i$ is given by (see Eq.~\ref{growth}):
\begin{equation}
    x'^{(i)}_0=\frac{x^{(i)}_0 e^{st}}{1+x^{(i)}_0(e^{st}-1)}=\frac{\lambda}{K\sum_{k=1}^D m_{ki}}=\frac{1}{K\sum_{k=1}^D m_{ki}}\left[1+st+\frac{(st)^2}{2}+o((st)^2)\right]\,,\label{growthi2}
\end{equation}
where we have again assumed $s>0$, $st\ll 1$, $K\gg 1$ and $Kst\gg 1$.

Upon the first dilution-migration, we sample on average $K m_{ij} x'^{(i)}_0$ mutants that go to deme $j$, using binomial sampling from the grown deme $i$. The relevant binomial law is $\mathcal{B}\left(N'_i,\,Km_{ij}x'^{(i)}_0/N'_i\right)$, where $N'_i$ is the size of deme $i$ after exponential growth. If $N'_i\gg 1$ and $Km_{ij}x'^{(i)}_0/N'_i\ll 1$ while $\lambda_{ij}=Km_{ij}x'^{(i)}_0$ is of order 1, we can approximate this by the Poisson distribution with mean 
\begin{equation}
    \lambda_{ij}=Km_{ij}x'^{(i)}_0=\lambda\frac{m_{ij}}{\sum_{k=1}^D m_{ki}}\equiv\lambda\mu_{ij}\,,
\end{equation}
where we used Eq.~\ref{growthi2} and introduced 
\begin{equation}
    \mu_{ij}=\frac{m_{ij}}{\sum_{k=1}^D m_{ki}}\,.
    \label{mus}
\end{equation}

We observe that the situation is exactly the same as above, except that $m_{ij}$ has been replaced by $\mu_{ij}$. All the statements above can be extended to the present case by replacing $m_{ij}$ by $\mu_{ij}$. In particular, we obtain equations on the mutant extinction probability similar to Eq.~\ref{genfsub}, namely:
\begin{equation}
    p_i=\exp\left[\lambda\sum_{j=1}^D \mu_{ij} \left(p_j-1\right)\right]\,.
    \label{genfsub2}
\end{equation}
Expanding in powers of $st$ then yields Eq.~\ref{pieq}, but with
\begin{align}
    a_i&=\sum_{j=1}^D \mu_{ij}a_j\,, \label{ai_mu}\\
    b_i&=\sum_{j=1}^D \mu_{ij}b_j-2a_i+a_i^2\,, \label{bi_mu}
    \\
    c_i&=\sum_{j=1}^D \mu_{ij}c_j-3 b_i + 3 a_i b_i - 3 a_i + 3 a_i^2 - 2 a_i^3 \label{ci_mu}\,.
\end{align}

\subsection{Strongly symmetric graphs}

\subsubsection{Some circulation graphs: clique and cycle}
\label{circ_example}
\paragraph{Clique.} In the clique, all demes are equivalent, and thus $a_i$ and $b_i$ do not depend on $i$. Furthermore, migration probabilities are all identical for $i\neq j$: say that $m_{ij}=m$ for all $i\neq j$ and $m_{ii}=\tilde{m}$ for all $i$. We observe that the two cases distinguished above coincide, because $\sum_{i=1}^Dm_{ij}=\sum_{j=1}^Dm_{ij}=\tilde{m}+(D-1)m$. Furthermore, in both cases, $\tilde{m}+(D-1)m=1$. In this context, Eq.~\ref{cd1} yields $a=a$ and Eq.~\ref{cd2} yields $b=b-2a+a^2$, and thus $a=2$, and we obtain an extinction probability $p=1-2st$, which corresponds to that of the well-mixed population in our serial dilution and exponential growth model (compared to the Wright-Fisher population considered above (see Eq.~\ref{WFfixBP}), $s$ is replaced by $st$). Thus, the clique structure has no effect on fixation probability in the regime addressed by the branching process formalism.

\paragraph{Cycle.} In the cycle, all demes are equivalent, and thus $a_i$ and $b_i$ do not depend on $i$. Let us introduce $m_{i,i+1}=m_C$, $m_{i,i-1}=m_A$ and $m_{ii}=\tilde{m}$ for all $i$ (with periodic boundary conditions). We have $\sum_{i=1}^Dm_{ij}=\sum_{j=1}^Dm_{ij}=\tilde{m}+m_A+m_C=1$, and thus, the two cases distinguished above coincide. In this context, as for the cycle, Eq.~\ref{cd1} yields $a=a$ and Eq.~\ref{cd2} yields $b=b-2a+a^2$, and thus $a=2$. Similarly, Eq.~\ref{cd3} yields $b=10/3$, and we obtain an extinction probability $p=1-2st+5/3(st)^2$: the cycle structure has no effect on fixation probability in the regime addressed by the branching process formalism.

\subsubsection{The star graph}
\label{star_BP_proba}
\paragraph{Star with $\sum_{i=1}^Dm_{ij}=1$ for all $j$ (all demes start at $K$ on average).} In the star, all leaves are equivalent, but the center is different. Denoting by $m_I$ migrations from a leaf to the center, by $m_O$ those from the center to a leaf, and by $\tilde{m}_C$ or $\tilde{m}_L$ the self-migrations for the center and the leaf, respectively, $\sum_{i=1}^Dm_{ij}=1$ yields
\begin{align}
    \tilde{m}_C&=1-(D-1)m_I\,,\\
    \tilde{m}_L&=1-m_O\,.
\end{align}
Eq.~\ref{cd1} becomes, considering the center and a leaf: 
\begin{align}
    a_C&=\tilde{m}_C a_C+(D-1)m_Oa_L=\left[1-(D-1)m_I\right]a_C+(D-1)m_Oa_L\,,\\
    a_L&=\tilde{m}_L a_L+m_I a_C=\left[1-m_O\right]a_L+m_I a_C\,,
\end{align}
which yields 
\begin{equation}
    a_L=\frac{m_I}{m_O}a_C=\alpha\, a_C\,,
\end{equation}
where we introduced the migration asymmetry $\alpha=m_I/m_O$.
Then, Eq.~\ref{cd2} becomes, considering the center and a leaf: 
\begin{align}
    b_C&=\left[1-(D-1)m_I\right]b_C+(D-1)m_Ob_L-2 a_C+a_C^2\,,\\
    b_L&=\left[1-m_O\right]b_L+m_I b_C-2 \alpha\,a_C+\alpha^2a_C^2\,,
\end{align}
which finally yields, assuming $a_C\neq 0$,
\begin{equation}
    a_C=2\,\frac{1+(D-1)\alpha}{1+(D-1)\alpha^2}\,\,\,\,\textrm{ and }\,\,\,\,a_L=\alpha\, a_C\,.
\end{equation}

Thus, the mutant fixation probabilities starting from one mutant at the bottleneck in the center or in the leaf of a star satisfying $\sum_{i=1}^Dm_{ij}=1$ for all $j$ are respectively:
\begin{equation}
    \rho_C=2st\frac{1+(D-1)\alpha}{1+(D-1)\alpha^2}\,\,\,\,\textrm{ and }\,\,\,\,\rho_L=2st\,\alpha\frac{1+(D-1)\alpha}{1+(D-1)\alpha^2}\,.
\end{equation}
We note that if $\alpha=1$, then $\rho_C=\rho_L=2st$, indicating no impact of structure on the fixation probability in this case. More generally, migration asymmetry $\alpha$ has a strong impact on these probabilities. Note however that only $\alpha$ enters our result, not $m_I$ and $m_O$ separately -- within the parameter regime where the branching process formalism holds. In the case where mutants are placed uniformly at random in any deme, the fixation probability is
\begin{equation}
    \rho_a=\frac{1}{D}\rho_C+\frac{D-1}{D}\rho_L=2st\frac{1}{D}\frac{\left[1+(D-1)\alpha\right]^2}{1+(D-1)\alpha^2}\,.
\end{equation}

\paragraph{Star with $\sum_{j=1}^Dm_{ij}=1$ for all $i$ (all demes contribute by $K$ on average).}
Using the same notations as above, $\sum_{j=1}^Dm_{ij}=1$ yields
\begin{align}
    \tilde{m}_C&=1-(D-1)m_O\,,\\
    \tilde{m}_L&=1-m_I\,.
\end{align}
Eq.~\ref{cd1} becomes, considering the center and a leaf: 
\begin{align}
    a_C&=\frac{\left[1-(D-1)m_O\right]a_C+(D-1)m_Oa_L}{1+(D-1)(m_I-m_O)}\,,\\
    a_L&=\frac{\left[1-m_I\right]a_L+m_I a_C}{1+m_O-m_I}\,,
\end{align}
which yields 
\begin{equation}
    a_L=\alpha\, a_C\,,
\end{equation}
exactly as above (recall that $\alpha=m_I/m_O$).
Then, Eq.~\ref{cd2} becomes, considering the center and a leaf: 
\begin{align}
    b_C&=\frac{\left[1-(D-1)m_O\right]b_C+(D-1)m_Ob_L}{1+(D-1)(m_I-m_O)}+a_C(a_C-2)\,,\\
    b_L&=\frac{\left[1-m_I\right]b_L+m_I b_C}{1+m_O-m_I}+\alpha\, a_C(\alpha\, a_C-2)\,,
\end{align}
which finally yields, assuming $a_C\neq 0$,
\begin{equation}
    a_C=2\,\frac{m_I-m_O+\frac{1}{D-1}+\alpha\left(1+m_O-m_I\right)}{m_I-m_O+\frac{1}{D-1}+\alpha^2\left(1+m_O-m_I\right)}\,\,\,\,\textrm{ and }\,\,\,\,a_L=\alpha\, a_C\,.
\end{equation}

Thus, the mutant fixation probabilities starting from one mutant at the bottleneck in the center or in the leaf of a star satisfying $\sum_{j=1}^Dm_{ij}=1$ for all $j$ are respectively:
\begin{equation}
    \rho_C=2st\,\frac{m_I-m_O+\frac{1}{D-1}+\alpha\left(1+m_O-m_I\right)}{m_I-m_O+\frac{1}{D-1}+\alpha^2\left(1+m_O-m_I\right)}\,\,\,\,\textrm{ and }\,\,\,\,\rho_L=\alpha\, \rho_C\,.
\end{equation}
We note again that if $\alpha=1$, then $\rho_C=\rho_L=2st$, indicating no impact of structure on the fixation probability in this case. More generally, migration asymmetry $\alpha$ enters these probabilities, but contrary to the previous case where migrations only entered through $\alpha$, here, $m_I$ and $m_O$ also matter separately.

\subsection{Circulation theorem: all circulation graphs have the same fixation probability}
\label{circ_thm}
Here, we extend the circulation theorem from Ref.~\cite{lieberman2005evolutionary} to our model, within the branching process approximation. Consider a metapopulation on a directed graph $G$ with a set of vertices $\mathbf{V}$, equipped with migration probabilities $m_{ij}\geq0$ from vertex $i$ to vertex $j$. $G$ is a circulation if and only if for all $i$, 
\begin{equation}
\sum_{j\in\mathbf{V}} m_{ij}=\sum_{j\in\mathbf{V}} m_{ji}\,,
\label{def_circ}
\end{equation}
which means that the sum of outgoing migrations from $i$ is equal to the sum of incoming migrations to $i$. We will restrict to connected graphs, such that there exists a path that connects each pair of vertices. In the regime of validity of the branching process approach, we aim to show that starting with a single mutant, all (connected) circulation graphs have the same probability of mutant extinction. Furthermore, this probability does not depend on the deme where the mutant started.

\subsubsection{Connected circulation graphs are strongly connected}
We start by showing that (connected) circulation graphs are strongly connected. A graph is strongly connected if each node can be reached from every other node. Note that this is stronger than connected because there needs to be a path from any node to any other in both directions.
\paragraph{Property of circulation graphs~\cite{lieberman2005evolutionary}.}
Consider a finite (connected) circulation graph $G$ with a set of nodes $\mathbf{V}$. Let $\mathbf{A}$ be a subset of nodes from a connected component of the graph. Then
\begin{align}
    \sum_{i\in \mathbf{A},\,j\in \mathbf{V}}m_{ij} &= \sum_{i\in \mathbf{A},\,j\in \mathbf{V}}m_{ji}\,,\,\,\textrm{i.e.\ ,} \\
    \sum_{i\in \mathbf{A},\,j\in \mathbf{V} \setminus \mathbf{A}}m_{ij} + \sum_{i\in \mathbf{A},\,j\in \mathbf{A}}m_{ij} &= \sum_{i\in \mathbf{A},\,j\in \mathbf{V}\setminus \mathbf{A}}m_{ji} + \sum_{i\in \mathbf{A},\,j\in \mathbf{A}}m_{ji}\,,
\end{align}
which results in 
\begin{equation}
    \sum_{i\in \mathbf{A},\,j\in \mathbf{V} \setminus \mathbf{A}}m_{ij} = \sum_{i\in \mathbf{A},\,j\in \mathbf{V} \setminus \mathbf{A}}m_{ji}\,. \label{circ-conservation}
\end{equation}
In other words, the sum of incoming migrations to $\mathbf{A}$ is equal to the sum of outgoing migrations from $\mathbf{A}$.
\paragraph{A transitive relation between strongly connected components.}
We define a transitive relation between two strongly components $\mathbf{A}, \mathbf{B}$. We say $\mathbf{A}$ is \emph{positively connected} to $\mathbf{B}$ and denote $\mathbf{A} \rightarrow_{+} \mathbf{B}$ if $$\sum_{i \in \mathbf{A}, j\in \mathbf{B}}m_{ij} > 0\,.$$
Given $m>1$ strongly connected components $\{\mathbf{A}_1,...,\mathbf{A}_m\}$, we say that they form a \emph{positive chain} if $\mathbf{A}_{k} \rightarrow_+ \mathbf{A}_{k+1}$ for $k \in \{ 1,...,m-1\}$.

\paragraph{Proof that $G$ is strongly connected.}
$G$ can be decomposed into an ensemble $\mathcal{S}$ of $n \geq 1$ maximal strongly connected components. If $n=1$, then $\mathcal{S}=G$ and the graph is strongly connected. Assume $n \geq 2$, let $\mathbf{A}_1 \in \mathcal{S}$. We aim to build a positive chain with distinct components of maximal length starting from $\mathbf{A}_1$.
\begin{itemize}[noitemsep,topsep=0pt]
    \item If $\sum_{i\in \mathbf{A}_1,\, j \in \mathbf{V}\setminus \mathbf{A}_1}m_{ij}=0$, then $m_{ij}=0$ for all $i\in \mathbf{A}_1,\,j \in \mathbf{V}\setminus \mathbf{A}_1$, because $m_{ij}\geq 0$. Eq.~\ref{circ-conservation} then further implies that $m_{ji}=0$ for all $i\in \mathbf{A}_1,\,j \in \mathbf{V}\setminus \mathbf{A}_1$. Thus, $\mathbf{A}_1$ is not connected to any other part of the graph, which is impossible, since the graph is connected. Therefore there exists another strongly connected component $\mathbf{A}_2 \in \mathcal{S}$ such that $\mathbf{A}_1 \rightarrow_+ \mathbf{A}_2$.
    \item For any $m \geq 2$, consider the set $\{\mathbf{A}_1, \mathbf{A}_2, ..., \mathbf{A}_m \} \in \mathcal{S}^m$ that completes $\{ \mathbf{A}_1, \mathbf{A}_2\}$ into a positive chain of $m$ distinct connecting components. The same argument as above implies that there exists  $\mathbf{A}_{m+1} \in \mathcal{S},\, \mathbf{A}_{m+1} \neq \mathbf{A}_{m}$ such that $\mathbf{A}_m \rightarrow_+ \mathbf{A}_{m+1}$. Moreover, $\mathbf{A}_{m+1} \neq \mathbf{A}_{k}\,,\,\, \forall k \in \{1,...,m\}$, otherwise $(\mathbf{A}_{k}, \mathbf{A}_{k+1},...,\mathbf{A}_{m}, \mathbf{A}_{m+1})$ would form one larger connected component, which is impossible since the $\mathbf{A}_i$ are maximal.
    \item By induction we build a positive chain of distinct maximal connected components of size $n$, where we recall that $n$ is the total number of maximal connected components in $G$. Then $\mathbf{A}_{n}$ should be positively connected to another element from $\mathcal{S}$, which is not in the chain. We reach a contradiction. 
\end{itemize}
In conclusion, $n=1$, and thus $\mathcal{S}=G$ and $G$ is strongly connected.
\subsubsection{Property of the migration matrix of a circulation graph} 
Let $M$ be the matrix with elements $\mu_{ij}=m_{ij}/\sum_{k} m_{ki}$. In the convention where $\forall i, \, \sum_{j} m_{ji}=1$, we have $\mu_{ij}=m_{ij}$. Furthermore, circulation graphs satisfy $\sum_{k} m_{ki}=\sum_k m_{ik}$, and thus, both in the convention where $\forall i, \, \sum_{j} m_{ji}=1$, and in the convention where $\forall i, \, \sum_{j} m_{ij}=1$, we have $\sum_{k} m_{ki}=\sum_k m_{ik}=1$. This yields $\mu_{ij}=m_{ij}$ also in the convention where $\forall i, \, \sum_{j} m_{ij}=1$.
Therefore, in both conventions, $M$ is the migration matrix, and all its lines and columns sum to 1. $M$ is also non-negative. 
In addition, as shown above, a circulation graph is strongly connected (i.e.\ there is a path in each direction between each pair of vertices of the graph), which entails that $M$ is irreducible (i.e.\ not similar via a permutation matrix to a block upper triangular matrix with more than one block). 

The Perron-Frobenius theorem for non-negative and irreducible matrices implies that $M$'s largest real eigenvalue $\rho$ is positive and satisfies
\begin{equation}
    \min_i \sum_j m_{ij} \leq \rho \leq \max_i \sum_j m_{ij} \,,
\end{equation}
and since both sums are equal to 1, $\rho=1$. The theorem also states that $\rho$ is a simple eigenvalue. Let $\mathcal{E}$ be the corresponding eigenspace of dimension 1. Since $\mathds{1}=(1,...,1)$ is an eigenvector associated to 1, it generates $\mathcal{E}$.

\subsubsection{Proof of the circulation theorem}
\paragraph{Goal of the computation.}
We are interested in the expansion of $p_i$, the probability of extinction starting from deme $i$, perturbatively in powers of $st$ that we rename $x$:
\begin{equation}
    p_i = 1 + \sum_{k=1}^\infty \alpha_k^{(i)}x^k\,.\label{exppi}
\end{equation}
Our goal is to show two points:
\begin{itemize}[noitemsep,topsep=0pt]
    \item For each order $k \geq 1$ and for all $i=1,...,D$, $\alpha_k^{(i)}=\alpha_k$, i.e. that the vector $(\alpha_k^{(i)})_{i=1,...,D} \in \mathcal{E}$. This entails that for circulations, $p_i$ does not depend on $i$: a mutant has the same probability of fixation, whatever the deme $i$ where it started.
    \item The equations satisfied by $\alpha_k$ for each $k \geq 1$ are the same for all circulations, i.e. they do not depend on the specific structure of the matrix $M$, provided that $\sum_{k} m_{ki}=\sum_k m_{ik}=1$ for all $i$. This entails that the fixation probability is the same for all circulations, which extends the circulation theorem~\cite{lieberman2005evolutionary} to our model in the regime of the branching process approximation.
\end{itemize}

\paragraph{Equation on $p_i$.}
We first expand Eq.~\ref{growthi} as
\begin{equation}
    \lambda=
 1 + \sum_{k=1}^\infty \lambda_k x^k\,,
\end{equation}
where $\lambda_1=1$ (see above).
We then write Eq.~\ref{genfsub} (or, equivalently here, Eq.~\ref{genfsub2}) using expansions in $x$:
\begin{align}
    p_i &= \exp \left[ \left( 1+ \sum_{l=1}^\infty \lambda_l x^l \right) \sum_{j=1}^D m_{ij} \left( \sum_{k=1}^\infty \alpha_k^{(j)}x^k \right)\right] \label{eq_pi}\\
     &= \exp \left[ \sum_{k=1}^\infty \beta_k^{(i)} x^k \right]\,,
\end{align}
where we introduced, for $k \geq 1$:
\begin{equation}
    \beta_k^{(i)} \equiv \sum_{j=1}^D m_{ij} \alpha_k^{(j)}+ \sum_{l=1}^{k-1} \sum_{j=1}^D m_{ij} \lambda_l \alpha_{k-l}^{(j)}\,. \label{beta_order}
\end{equation}
Expanding the exponential and replacing $p_i$ by its expression in Eq.~\ref{exppi} yields
\begin{equation}
    1 + \sum_{k=1}^\infty \alpha_k^{(i)}x^k = 1 + \sum_{n=1}^\infty \dfrac{1}{n!} \left( \sum_{k=1}^\infty \beta_k^{(i)} x^k\right)^n\,.
\end{equation}
Matching each order yields for $k \geq 1$:
\begin{equation}
    \alpha_k^{(i)}=\sum_{n=1}^k \dfrac{1}{n!} \sum_{\substack{k_1,...,k_n\geq 1\\ k_1 +...+k_n = k}} \beta_{k_1}^{(i)}...\beta_{k_n}^{(i)}\,. \label{match_order}
\end{equation}

\paragraph{Induction on the order.}
\label{Jordan_decomposition}
\begin{itemize}
    \item \textbf{Order 1:}
\end{itemize}
At order 1, Eq.~\ref{match_order} reduces to
\begin{equation}
    \alpha_1^{(i)}=\sum_{j=1}^D m_{ij} \alpha_1^{(j)}\,.
\end{equation}
In other words, the vector $(\alpha_1^{(1)},...,\alpha_1^{(D)})$ is an eigenvector of $M$ with eigenvalue 1. Thus, all its components are equal to a constant $\alpha_1$. \\
At order 2, using $\lambda_1=1$, Eq.~\ref{match_order} reads for $i=1,...,D$:
\begin{equation}
\alpha_2^{(i)}= \sum_{j=1}^D m_{ij} \alpha_2^{(j)} + \alpha_1 +\frac{1}{2} \alpha_1^2\,.
\end{equation}
 In vectorial terms, defining $A_2=(\alpha_2^{(1)},..., \alpha_2^{(D)})$, this equation reads
\begin{equation}
    A_2 = M A_2 + \left(\alpha_1 + \frac{1}{2}\alpha_1^2 \right) \mathds{1}\,.
\end{equation}
Using the Jordan decomposition, we complete $\mathcal{E}$ into a complex basis in which $M$ has a Jordan normal form. The complementary space of $\mathcal{E}$, that we call $\mathcal{E}^{\perp}$, is thus stable under the action of the matrix $M$. We decompose $A_2$ on this basis as $A_2 = A_{\parallel} + A_{\perp}$, with $A_{\parallel}$ in $\mathcal{E}$ and $A_{\perp}$ in $\mathcal{E}_{\perp}$. Then
\begin{equation}
    A_{\parallel} + A_{\perp} = M A_{\parallel} + M A_{\perp} + \left(\alpha_1 + \frac{1}{2}\alpha_1^2 \right) \mathds{1}\,,
\end{equation}
yielding, as $A_{\parallel} = M A_{\parallel}$,
\begin{equation}
    A_{\perp}- M A_{\perp}  = \left(\alpha_1 + \frac{1}{2}\alpha_1^2 \right) \mathds{1}\,.
\end{equation}
Therefore, $A_{\perp}- M A_{\perp} \in \mathcal{E} \cap \mathcal{E}_{\perp} = \{ 0 \}$, which entails $A_{\perp}\in \mathcal{E} \cap \mathcal{E}_{\perp} = \{ 0 \}$. This determines the value of $\alpha_1$ through
\begin{equation}
    2 \alpha_1 + \alpha_1^2 = 0\,,
\end{equation}
which yields $\alpha_1 = -2$ for all circulations, consistent with our results obtained for the clique and the cycle above.

\begin{itemize}
    \item \textbf{From order $k-1$ to order $k$:}
\end{itemize}
We assume that there exists $k \geq 2$, such that for all $l \leq k-1$, $(\alpha_{l}^{(1)},...,\alpha_{l}^{(D)}) = \alpha_l \mathds{1} \in \mathcal{E}$, which directly entails $(\beta_{l}^{(1)},...,\beta_{l}^{(D)}) \in \mathcal{E}$. We also assume that the values of $\alpha_l$ for $l \leq k-1$ are the same for all circulations. \\\\
From Eq.~\ref{match_order} written for order $k$ and for all $i=1,...,D$, we detail
\begin{align}
    \alpha_k^{(i)}&=\beta_k^{(i)}+\sum_{n=2}^{k} \dfrac{1}{n!} \sum_{\substack{k_1,...,k_n \geq 1\\ k_1 +...+k_n = k}} \beta_{k_1}^{(i)}...\beta_{k_n}^{(i)} \\
    &=\sum_{j=1}^D m_{ij} \alpha_k^{(j)}+ \sum_{l=1}^{k-1} \sum_{j=1}^D m_{ij} \lambda_l \alpha_{k-l}^{(j)}  +  \sum_{n=2}^{k} \dfrac{1}{n!} \sum_{\substack{k_1,...,k_n \geq 1\\ k_1 +...+k_n = k}} \beta_{k_1}^{(i)}...\beta_{k_n}^{(i)}\,.
\end{align}
In the last term, consider an ensemble of $n$ integers $k_1,...,k_n \geq 1$, such that $k_1 +...+k_n = k$. Since $n \geq 2$, we necessarily have $k_1,...,k_n \leq k-1$. According to the induction hypothesis, none of the $\beta_l^{(i)}$ in this sum depends on the index $(i)$. The second term also contains $(\alpha_l^{(j)})$s with $l \leq k-1$, which thus do not depend on the index $(j)$. We can rewrite this equation for $i=1,...,D$ as
\begin{equation}
    \alpha_k^{(i)} = \sum_{j=1}^D m_{ij} \alpha_k^{(j)} + \gamma_k^{(i)}\,, \label{alphak_simple}
\end{equation}
where we introduced
\begin{align}
 \gamma_k &= \sum_{l=1}^{k-1} \lambda_l \sum_{j=1}^D m_{ij} \alpha_{k-l}^{(j)}  +  \sum_{n=2}^{k} \dfrac{1}{n!} \sum_{\substack{k_1,...,k_n \geq 1\\ k_1 +...+k_n = k}} \beta_{k_1}^{(i)}...\beta_{k_n}^{(i)}\nonumber\\
 &=\sum_{l=1}^{k-1} \lambda_l \alpha_{k-l}\sum_{j=1}^D m_{ij}   +  \sum_{n=2}^{k} \dfrac{1}{n!} \sum_{\substack{k_1,...,k_n \geq 1\\ k_1 +...+k_n = k}} \beta_{k_1}...\beta_{k_n}\nonumber\\
 &=\sum_{l=1}^{k-1} \lambda_l \alpha_{k-l}   +  \sum_{n=2}^{k} \dfrac{1}{n!} \sum_{\substack{k_1,...,k_n \geq 1\\ k_1 +...+k_n = k}} \beta_{k_1}...\beta_{k_n}\,,
\end{align}
where we employed $\sum_{j=1}^D m_{ij}=1$, which holds for circulations in both conventions. Importantly,  $\gamma_k$ only involves lower-order terms and does not depend on $i$. In vectorial form, defining $A_k=(\alpha_k^{(1)},..., \alpha_k^{(D)})$, Eq.~\ref{alphak_simple} reads
\begin{equation}
    A_k = M A_k + \gamma_k \mathds{1}\,.
\end{equation}
We apply the same argument as above for order 2, using the Jordan decomposition of $M$ into a basis $\mathcal{E} \bigoplus \mathcal{E}^{\perp}$.
It entails that $A_k \in \mathcal{E}$, i.e. all its components $\alpha_k^{(1)},..., \alpha_k^{(D)}$ are equal to a constant $\alpha_k$, and $\gamma_k=0$.
Considering Eq.~\ref{alphak_simple} at order $k+1$ yields $\gamma_{k+1}=0$, i.e. 
\begin{equation}
\sum_{l=1}^{k} \lambda_l \alpha_{k+1-l}  =-  \sum_{n=2}^{k+1} \dfrac{1}{n!} \sum_{\substack{k_1,...,k_n \geq 1\\ k_1 +...+k_n = k+1}} \beta_{k_1}...\beta_{k_n}\,.
\end{equation}
This yields a function of the $\alpha_k$ that does not depend on the $m_{ij}$.

 Therefore the value of $\alpha_k$ is the same for all circulations, which concludes the proof by induction.

 \subsubsection{Circulation graphs have the same fixation probability as the well-mixed population}
We have shown that for circulation graphs, the extinction probability $p_i$ of one mutant does not depend on the initial deme $i$. Let us denote it by $\tilde{p}$. Applying Eq.~\ref{genfsub} to a circulation graph (which satisfies $\sum_i m_{ij}=\sum_i m_{ji}=1$) yields
\begin{equation}
    \tilde{p}=\exp[\lambda (\tilde{p}-1)]\,.
\end{equation}
This is the same equation as Eq.~\ref{genf}, which holds for the extinction probability $p$ of one mutant in a well-mixed population. This equation admits a unique solution apart from $1$, which entails $p = \tilde{p}$. 

Note that rewriting Eq.~\ref{exppi} as
\begin{equation}
    \tilde{p}= 1 + \sum_{k=1}^\infty \alpha_k x^k \,,
\end{equation}
and applying Eq.~\ref{eq_pi} gives the same result.

Therefore, within the branching process description, all circulation graphs have exactly the same fixation probability as the well-mixed population.

\subsection{Circulations have the same average extinction time as a well-mixed population}
\label{exttimecirc}

The average extinction time in a structured population is given by Eq.~\ref{tex}, and that in a well-mixed population is given by a similar formula involving the well-mixed generating function $g$ (see Eq.~\ref{genf}).

Consider a circulation initialized with a single mutant in one deme. In the branching process approximation, the circulation has generating function $\mathbf{h}$, with components $(h_1,...,h_D)$. Let us show by induction that for all $n\geq 1$, the $n$-th iterate of $\mathbf{h}$ applied to vector $\mathbf{0}=(0,...,0)$ has all its components equal to $g^{(n)}(0)$, which is the $n$-th iterate of the well-mixed generating function $g$ (see Eq.~\ref{genf}) applied to 0.
\begin{itemize}
    \item \textbf{Order 1:}
\end{itemize}
For all $i = 1,...,D$, using Eq.~\ref{gen_Poisson} and exploiting the fact that $\sum_j m_{ij}=\sum_j m_{ji}=1$ for circulations gives:
\begin{equation}
    h_i(\mathbf{0})=\exp \left( - \lambda \sum_j m_{ij}\right) = \exp(-\lambda)=g(0).
\end{equation}
\begin{itemize}
    \item \textbf{From order $k-1$ to order $k$:}
\end{itemize}
Assume there exists $k> 1$ such that $\mathbf{h}^{(k-1)}(\mathbf{0})=\left(g^{(k-1)}(0),\dots,g^{(k-1)}(0)\right)$. Then for all $i = 1,\dots,D$:
\begin{equation}
    (\mathbf{h}^{(k)}(\mathbf{0}))_i = h_i (\mathbf{h}^{(k-1)}(\mathbf{0})) = \exp \left( - \lambda \sum_j m_{ij} g^{(k-1)}(0) \right) = \exp (- \lambda g^{(k-1)}(0)) = g^{(k)}(0).
\end{equation}
This concludes the proof by induction.
Using this equality in Eq.~\ref{tex}, we find that the average extinction time for a circulation starting from a mutant in any deme $i$ is the same as the average extinction time in the well-mixed population. This holds in the regime of frequent migrations where the sampling is well-described by the Poisson approximation.

\subsection{All graphs that are not circulations strictly suppress selection}
\label{strict_supp}
Consider a mutant introduced uniformly at random in a graph at the bottleneck, i.e. with probability proportional to deme bottleneck sizes. (If all demes start at bottleneck size $K$ on average, as in the main text, this means that a mutant is introduced in an initial deme picked uniformly at random.) Here, we will show that for any connected graph that is not a circulation, the mutant fixation probability is strictly smaller than $2st$, the fixation probability in well-mixed populations and circulations. This means that all graphs that are not circulations strictly suppress selection. 

For this, we focus on first order terms of the expansions of the extinction probabilities $p_1,...,p_D$ in powers of $st$, namely $a_1,...,a_D$. For circulation graphs, we have already proven that $a_i=2 \,\, \forall i$. We show that the average of $a_1,...,a_D$ weighted by deme bottleneck sizes is always smaller than or equal to 2.

\subsubsection{When all demes start at $K$ on average}
\label{eigval1}
In this convention, $\sum_{i=1}^Dm_{ij}=1$ for all $j$. Here, a mutant is introduced in an initial deme picked uniformly at random, at the bottleneck. Then, the average extinction probability is obtained to first order in $st$ by computing the arithmetic mean of the $a_i$.

We start from Eqs.~\ref{cd1} and \ref{cd2}. Eq.~\ref{cd1} shows that $(a_1,...,a_D)$ is an eigenvector associated to the eigenvalue 1. Such an eigenvector exists, as the migration matrix is a column-stochastic matrix (it has non-negative entries with columns summing to 1), which entails that 1 is its largest eigenvalue.  In addition, $a_i\geq0$ for all $i$ since $p_i < 1$.  

We next consider Eq.~\ref{cd2} to specify the values of $a_1,...,a_D$.  Summing this equation over $i$, we write:
\begin{align*}
    \sum_{i=1}^D b_i &= \sum_{i,j} m_{ij} b_j - 2 \sum_{i=1}^D a_i + \sum_{i=1}^D a_i^2\,,\,\,\textrm{i.e.\ ,} \\
    \sum_{i=1}^D b_i &= \sum_{j=1}^D b_j - 2 \sum_{i=1}^D a_i + \sum_{i=1}^D a_i^2\,,\,\,\textrm{i.e.\ ,} \\
    2 \sum_{i=1}^D a_i &= \sum_{i=1}^D a_i^2 \,.
\end{align*}
We now apply Cauchy-Schwarz's inequality to the vectors $A=(a_1,\dots,a_D)$ and $\mathds{1}=(1,\dots,1)$, yielding:
\begin{equation}
    \left(\sum_{i=1}^D a_i\right)^2 \leq D \sum_{i=1}^D a_i^2 =  2 D \sum_{i=1}^D a_i\,,
\end{equation}
and therefore 
\begin{equation}
\dfrac{1}{D} \sum_{i=1}^D a_i \leq 2\,. \label{supp_ineq}
\end{equation} 

Moreover, Eq.~\ref{supp_ineq} is an equality if and only if vectors $A$ and $\mathds{1}$ are colinear, in which case all $a_i$s are equal. Eq.~\ref{cd1} then yields $\sum_j m_{ij}=1$ for all $i$, hence the graph is a circulation.
Thus, all graphs that are not circulations strictly suppress selection.

\subsubsection{When all demes contribute by $K$ on average}
In this convention, $\sum_{i=1}^Dm_{ji}=1$ for all $j$. 
The proof in this case is very similar, but we start from Eqs.~\ref{ai_mu} and \ref{bi_mu}. In addition, here the bottleneck size of deme $i$ is $KC_i$, where $C_i = \sum_j m_{ji}$ is the sum of incoming migrations to deme $i$ (none of them are equal to 0, since no deme is disconnected). Thus, the average extinction probability is obtained to first order in $st$ by computing the average of the values of $a_i$  weighted by $C_i/D$.

First, let us make sure that matrix $\mu$, with elements $\mu_{ij} = m_{ij}/C_i$, has 1 as eigenvalue, thereby ensuring that Eq.~\ref{ai_mu} has a non-trivial solution. For this, we show that its spectrum is the same as that of another matrix whose columns sum to 1. Indeed, let $\eta$ be an eigenvalue of $\mu$, and $y$ an associated eigenvector. We then have $\forall i \in \{1,...,D \}$:
\begin{align}
    \eta\, y_i &= \sum_{j=1}^D \mu_{ij} y_j\,,\,\,\textrm{i.e.\ ,}\\
   \eta\, y_i &=  \sum_{j=1}^D\dfrac{m_{ij}}{C_i} y_j\,,\,\,\textrm{i.e.\ ,}\\
   \eta\, y_i C_i &=  \sum_{j=1}^D \dfrac{m_{ij}}{C_j} y_j C_j.
\end{align}
Thus, $\eta$ is also an eigenvalue of matrix $\tilde{\mu}$ defined by $\tilde{\mu}_{ij}=m_{ij}/C_j$. This entails that $\mu$ and $\tilde{\mu}$ share the same spectrum. But $\tilde{\mu}$ is non-negative with columns summing to 1, and therefore 1 is indeed an eigenvalue (the largest). 

We now want to characterize eigenvector $(a_1,...,a_D)$ starting from Eq.~\ref{bi_mu}:
\begin{align}
    b_i &= \sum_{j=1}^D \frac{m_{ij}}{C_i} b_j - 2 a_i + a_i^2 \,,\,\,\textrm{i.e.\ ,}\\
    C_i b_i &= \sum_{j=1}^D \frac{m_{ij}}{C_j} C_j b_j - 2 C_i a_i + C_i a_i^2\,.
\end{align}
Summing over $i$ on both sides, and recalling that $\sum_i m_{ij} = C_j$, we obtain
\begin{align}
    \sum_{i=1}^D  C_i b_i &= \sum_{j=1}^D C_j b_j - 2 \sum_{i=1}^D C_i a_i + \sum_{i=1}^D C_i a_i^2 \,,\,\,\textrm{i.e.\ ,}\\
    2 \sum_{i=1}^D C_i a_i &= \sum_{i=1}^D C_i a_i^2\,.
\end{align}
Applying Cauchy-Schwarz's inequality to the vectors $(\sqrt{C_1},\dots,\sqrt{C_D})$ and $(\sqrt{C_1}a_1,\dots,\sqrt{C_D}a_D)$ yields
\begin{equation}
\left(\sum_{i=1}^D C_i a_i\right)^2 \leq \sum_{i=1}^D C_i  \sum_{j=1}^D C_j a_j^2=2\sum_{i=1}^D C_i  \sum_{j=1}^D C_j a_j\,.
\end{equation}
Since $\sum_{i=1}^D C_i = D$, we obtain
\begin{equation}
    \left(\sum_{i=1}^D C_i a_i\right)^2 \leq 2D \sum_{i=1}^D C_i a_i\,,
\end{equation}
which yields the desired result:
\begin{equation}
\sum_{i=1}^D \frac{C_i}{D} a_i \leq 2\,. \label{supp_ineq_mu}
\end{equation}

Eq.~\ref{supp_ineq_mu} is an equality if and only if $(\sqrt{C_1},\dots,\sqrt{C_D})$ and $(\sqrt{C_1}a_1,\dots,\sqrt{C_D}a_D)$ are colinear. In that case, all $a_i$s are equal to 2, and Eq.~\ref{ai_mu} yields $\sum_j \mu_{ij}=1$ for all $i$, hence the graph is a circulation. 

To summarize, we have shown that in the branching process approach, and if migration probabilities are all of order 1 (which allows to use the Poisson distribution), any graph that is not a circulation strictly suppresses natural selection to first order in $st$. In contrast, circulation graphs have the same fixation probability as a well-mixed population.

\section{Structured populations on graphs with serial dilutions: branching process with binomial sampling}
\label{binSampl}

\subsection{Motivation}

Recall that upon the first dilution-migration step, we sample on average $K m_{ij} x'^{(i)}_0$ mutants that go to deme $j$ using binomial sampling from the grown deme $i$. The Poisson approximation made above requires $\lambda_{ij}=Km_{ij}x'^{(i)}_0$ to be of order 1. However, this does not hold if $m_{ij}$ is of order $st$ or smaller. Moreover, the identification of terms in expansions in powers of $st$ that we performed above to obtain extinction probabilities has to be handled differently for these small $m_{ij}$. To treat the case of rarer migrations, of order $st$ or $(st)^2$, we thus need to generalize our treatment and to go back to the complete binomial law. However, we will remain within the multi-type branching process approximation, assuming $s>0$, $st\ll 1$, $K\gg 1$ and $Kst\gg 1$. 

First, we will treat the convention $\sum_i m_{ij} = 1$, i.e.\ all demes start at $K$ on average. The corresponding results are shown in the main text. Then, we will discuss how these calculations can be extended to the other convention $\sum_j m_{ij} = 1$, where all demes contribute by $K$ on average, and show the corresponding results.

\subsection{When all demes start at $K$ on average}
\label{ex_proba}
\subsubsection{Generating function}

The generating function $\mathbf{f}(\mathbf{x})$ is given by Eq.~\ref{genfctgal}. Starting from one mutant in deme $i$, with deme size equal to $N'_i$ and mutant ratio $x_0'^{(i)}$ after growth, the number of mutants sent from $i$ to $j$ is sampled from  the binomial law $\mathcal{B}\left(N'_i, K x_0'^{(i)} m_{ij}/N'_i\right)$.
    Therefore
    \begin{equation}
        \phi_i(r_1,...,r_D)= \prod_{j=1}^D \binom{N'_i}{r_j} \left( \dfrac{K}{N'_i} x_0'^{(i)} m_{ij} \right)^{r_j} \left( 1-\dfrac{K}{N'_i} x_0'^{(i)} m_{ij} \right)^{N'_i-r_j}.
    \end{equation}
    The $i$-th component of the generating function now reads
\begin{equation}
    f_i(\mathbf{x})=\prod_{j=1}^D \left( 1 - (1-x_j) \dfrac{\lambda}{N'_i}m_{ij}\right)^{N'_i}, \label{generating_bin}
\end{equation}
where $\lambda$ is given by Eq.~\ref{lamWF}. 

One can consider a well-mixed population with sampling performed exactly as in our structured populations on graphs (note that this is slightly different from the case considered in Section~\ref{wmpop} where bottleneck size was exactly $K$). This is useful as reference. Then, the (single-type) generating function reads
\begin{equation}
    f(x)= \left( 1 - (1-x)\dfrac{\lambda}{N'}\right)^{N'} \,,
    \label{fwm}
\end{equation}
where $N'$ is the size of the population after growth.

\subsubsection{Extinction probabilities: general expansion}

Let $p_i$ be the extinction probability starting from one mutant in deme $i$. Under the assumptions $s>0$, $st\ll 1$, $K\gg 1$ and $Kst\gg 1$, let us expand $p_i$ perturbatively in powers of $st$:
\begin{equation}
    p_i=1-a_ist+\frac{b_i}{2}(st)^2-\frac{c_i}{6}(st)^3+o((st)^3)\,. \label{pieqb}
\end{equation}
The vector of extinction probabilities $\mathbf{p}=(p_i,...,p_D)$ is a fixed point of the generating function. For each component $i$, we expand pertubatively in powers of $st$ the equality
\begin{equation}
    p_i = f_i (\mathbf{p})\,.
\end{equation}
Note that $(1-p_j) \lambda m_{ij}$ is at most of order $st$ (if $m_{ij}$ is of order one). Furthermore, since $N'_i > K$ for all $i$, and $st\gg 1/K$, we have $st \gg 1/N'_i$. We can then expand the generating function components as
\begin{align}
    f_i(\mathbf{p}) &= \prod_{j=1}^D \left[ 1 - (1-p_j) \lambda m_{ij} + \dfrac{N'_i (N'_i-1)}{2} \left( (1-p_j) \dfrac{\lambda}{N'_i} m_{ij} \right)^2 \right.\nonumber\\ &\phantom{hellohi}\left. -\, \dfrac{N'_i (N'_i-1) (N'_i-2)}{6} \left( (1-p_j) \dfrac{\lambda}{N'_i} m_{ij} \right)^3 + o((st)^3)\right] \nonumber\\
    &= \prod_{j=1}^D \left[ 1 - (1-p_j) \lambda m_{ij} + \dfrac{1}{2}\left( (1-p_j) \lambda m_{ij} \right)^2  - \dfrac{1}{6} \left( (1-p_j) \lambda m_{ij} \right)^3 + o((st)^3)\right]. \label{gen_expansion}
\end{align}
To further expand Eq.~\ref{gen_expansion}, we need to specify how migration probabilities scale with $st$. Below, we consider different regimes.

Note that the $i$-th component of the generating function in Eq.~\ref{generating_bin} can be written as
\begin{equation}
    f_i(\mathbf{x})= \prod_{j=1}^D \exp \left[ N'_i \log \left( 1 - (1-x_j)\dfrac{\lambda}{N'_i}\right)\right].
\end{equation}
If we take the limit $N'_i \rightarrow \infty$, then for any $k$, $(st)^k \ll 1/N'_i$.
Applying $f_i$ to $\mathbf{p}$, expanding the logarithm in the powers of $st$ and discarding terms of order $1/N'_i$ then yields
\begin{equation}
    f_i(\mathbf{p})=\exp \left[ \lambda \sum_{j=1}^D m_{ij} (p_j -1)\right] \,,
\end{equation}
which is the same generating function as the one we obtained with the Poisson approximation in Eq.~\ref{gen_Poisson}. This is consistent, since the Poisson approximation requires $N'_i \rightarrow \infty$. However, for large but finite populations, one should assume that $N'_i$ is finite, meaning that $(st)^k \ll 1/N'_i$ will break down beyond some $k$. For this reason, we present a detailed analysis of the binomial case, restricting to the first terms of the expansion in $st$, detailed in Eq.~\ref{gen_expansion}.

\subsubsection{Well-mixed population}
\label{ex_proba_wm}
 For a well-mixed population, the probability of extinction $p = 1 -a st + b (st)^2/2 + o ((st)^2)$ of the mutant satisfies the equation $p = f (p)$ with $f$ defined in Eq.~\ref{fwm}. Expanding this equation in powers of $st$ and identifying terms yields $a = 2$ and $b=10/3$. This result coincides with the one obtained in Section~\ref{wmpop}, despite the minor difference in sampling. Note that the Poisson approximation holds in the well-mixed case.

\subsubsection{When all migration probabilities are of order 1}
Assuming that all migration probabilities are of order 1, for all $i$, $f_i (\mathbf{p})$ in Eq.~\ref{gen_expansion} can be expanded up to third order in $st$ as:
\begin{align}
    f_i (\mathbf{p}) = \prod_{j=1}^D &\left[ 1 -  m_{ij} a_j st +\frac{m_{ij}}{2} \left( b_j-2a_j+m_{ij}a_j^2  \right) (st)^2 \right. \\ &\left. -\,\frac{m_{ij}}{6} \left( c_j-3b_j+3a_j+3m_{ij}a_jb_j-6m_{ij}a_j^2+m_{ij}^2a_j^3\right)(st)^3\right] . \nonumber
\end{align}

Using the expansion of $p_i$ in Eq.~\ref{pieqb}, identifying the first and second order terms in $st$ in the equality $p_i = f_i (\mathbf{p})$ yields:
\begin{align}
    a_i &= \sum_j m_{ij} a_j\,, \\
    b_i &= \sum_j m_{ij} b_j - 2 a_i + a_i^2\,.
\end{align}
These equations match those we obtained with the Poisson law. Again, thanks to the Cauchy-Schwartz inequality, we show that $\sum_i a_i/D \leq 2$, and the case of equality corresponds to $a_i = 2$ for all $i$. Then, $\sum_j m_{ij}=1$ for all $j$, therefore the graph is a circulation. Hence, suppression is strict to first order in $st$ for all graphs that are not circulations and where migration probabilities are all of order 1.

Identifying the third order terms in $st$ in the equality $p_i = f_i (\mathbf{p})$ further yields
\begin{equation}
    c_i=\sum_jm_{ij}c_j-3b_i+3a_ib_i-3a_i+3a_i^2-2a_i^3\,.
\end{equation}
If the graph is a circulation, then by the same argument detailed in~\ref{Jordan_decomposition}, using the Jordan normal form of the migration matrix, we obtain $b_i=10/3$ for all $i$. This is consistent with the second order coefficient obtained using the Poisson approximation for circulation graphs in~\ref{circ_example}, and coincides with the well-mixed population result.

\subsubsection{When exchanges between different demes are of order $st$}
This time, assume that migration probabilities between different demes are of order $st$, while self-loops can have a dominant term of order 1.
We write
\begin{align}
    m_{ij} &= m_{ij}^{(1)}st \,\,\,\,\mathrm{if}\,\, i\neq j\,,\\
    m_{ii} &= 1+m_{ii}^{(1)}st\,.
\end{align}
Expanding Eq.~\ref{gen_expansion} yields
\begin{align}
    a_i &= - \sum_j a_j m_{ij}^{(1)} +\dfrac{a_i^2}{2}\,, \\
0&=-3a_i+3a_i^2-2a_i^3-3b_i+3a_ib_i-3\sum_j b_j m_{ij}^{(1)}\,.
\end{align}

From $\sum_i m_{ij} = 1$, we deduce $\sum_im_{ij}^{(1)} = 0$. Summing the two equations above on $i$, we can again apply the Cauchy-Schwartz inequality to get $\sum_i a_i/D \leq 2$. Note that the case of equality gives $a_i=2$ for all $i$, and then the second equation simplifies to $\sum_i b_i/D =10/3$. In that case, the second order term is the same starting from any deme ($b_i = 10/3$ for all $i$) if the graph is a circulation.

\subsubsection{When exchanges between different demes are of order $(st)^2$}
This time, we write
\begin{align}
    m_{ij} &= m_{ij}^{(2)}(st)^2 \,\,\,\,\mathrm{if}\,\, i\neq j\,,\\
    m_{ii} &= 1 + m_{ii}^{(2)}(st)^2\,.
\end{align}
Expanding Eq.~\ref{gen_expansion} yields
\begin{align}
    a_i &= 2\,,  \\
    0&=3a_i-6a_i^2+a_i^3-3b_i+3a_ib_i+6\sum_j a_j m_{ij}^{(2)}\,,
\end{align}
and the second term becomes
\begin{equation}
    b_i = \dfrac{10}{3}-4\sum_j m_{ij}^{(2)}. \label{bi_gen}
\end{equation}
Summing on $i$ and using $\sum_i m_{ij}^{(2)}=0$ yields
\begin{equation}
    \dfrac{1}{D}\sum_i b_i = \dfrac{10}{3} \,.
\end{equation}
Thus, in this case, the fixation probability is the same as in the well-mixed case up to first order. Furthermore, the fixation probability averaged over all starting demes is the same as in the well-mixed case up to second order. Note that Eq.~\ref{bi_gen} entails that the second order term is the same starting from any deme ($b_i = 10/3$ for all $i$) if and only if the graph is a circulation.

\subsubsection{Application to the star graph}

\begin{figure}[htbp]
\begin{center}
\includegraphics[scale=0.52]{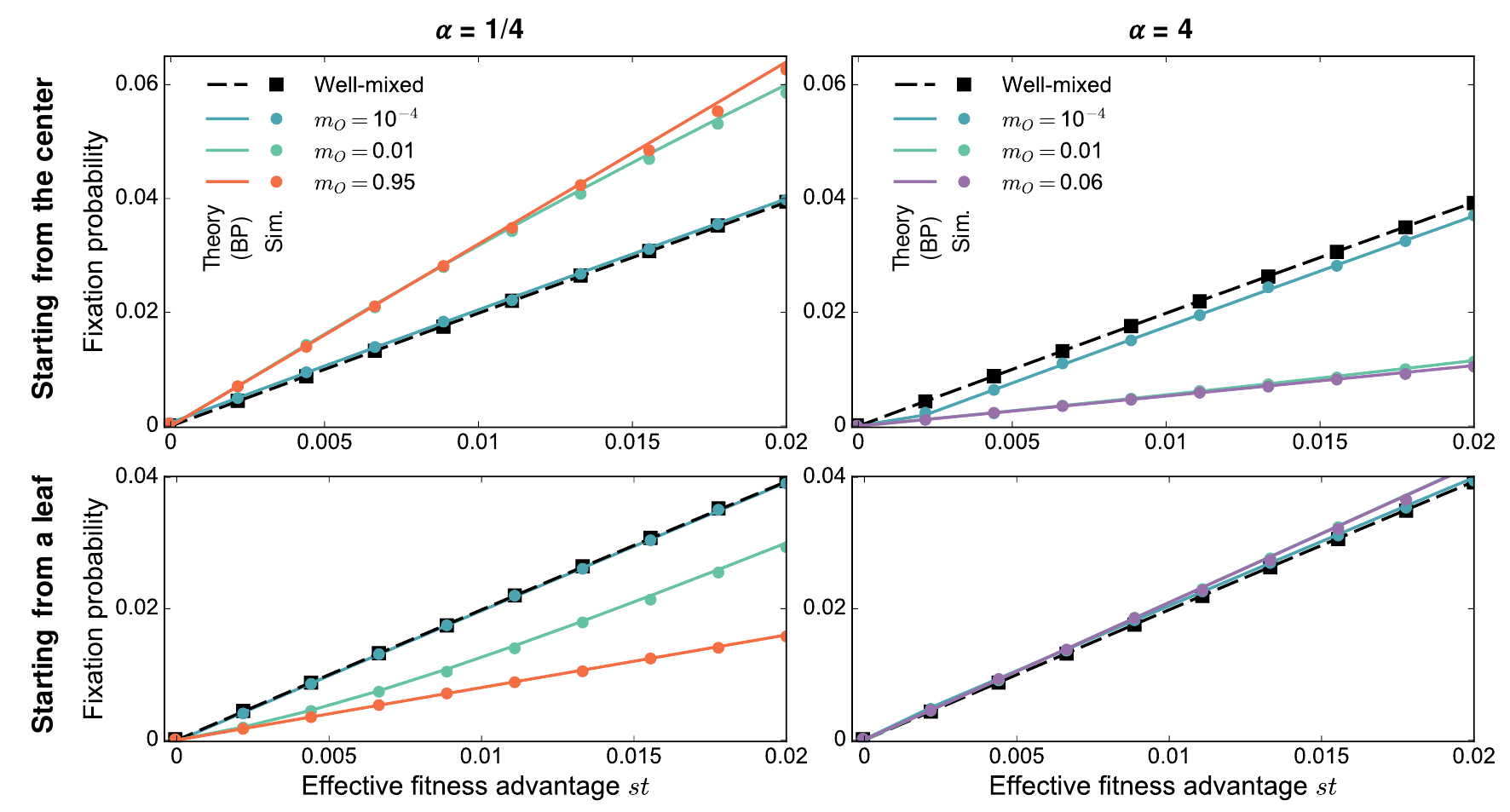}
\caption{\textbf{Mutant fixation in the star starting from the center or from a leaf.} Mutant fixation probability is plotted versus the effective fitness advantage $st$ of the mutant. We consider a star  with $D=5$ demes, and $K=1000$ individuals per deme on average at the bottleneck, for migration asymmetries $\alpha=1/4$ (left) and $\alpha=4$ (right). The growth phase duration is $t=5$.  We start with one mutant of fitness $f_M=1+s$ placed in the center (top) or in one of the leaves (bottom), all other individuals being wild-types with fitness $f_W=1$. Markers are simulation results (``Sim.'') averaged on 1 million realizations.  Lines are theoretical predictions obtained through a branching process (``BP'') approach. The well-mixed case is shown for comparison, with simulations performed for a well-mixed population with $KD=5000$ individuals at the bottleneck, initialized with one mutant.}
\label{star_fprobas_BP_centerleaf}
\end{center}
\end{figure}

We apply the results from the three previous subsections to the star graph (see Fig.~\ref{model_structures}), with $D=5$ demes, $K=1000$ individuals per deme on average at the bottlenecks, and a migration asymmetry $\alpha= m_I / m_O$. We compute the first order term $a_C$ (resp. $a_L$, see Section~\ref{star_BP_proba}) of the fixation probability of one mutant starting in the center (resp. starting in a leaf), in different regimes of migration probabilities. Fig.~\ref{star_fprobas_BP_centerleaf} shows that the branching process predictions agree very well our stochastic simulations. In particular, we see that starting from the center for $\alpha>1$ amplifies natural selection for frequent migrations, while starting from the leaf suppresses selection. The opposite holds for $\alpha<1$: starting from a leaf amplifies selection while starting from the center suppresses it. Averaging over the initial position of the mutant always results in suppression, as shown in Fig.~\ref{star_fprobas_etimes_BP}.

\newpage

\subsubsection{Numerical results on fixation dynamics}
\label{FixDyn}

While we obtained results on fixation probabilities and extinction times, another important question regards the dynamics of fixation. Does spatial structure impact how fast fixation occurs? We addressed this question in numerical simulations. 

\begin{figure}[h!]
\begin{center}
\includegraphics[width=0.65\textwidth]{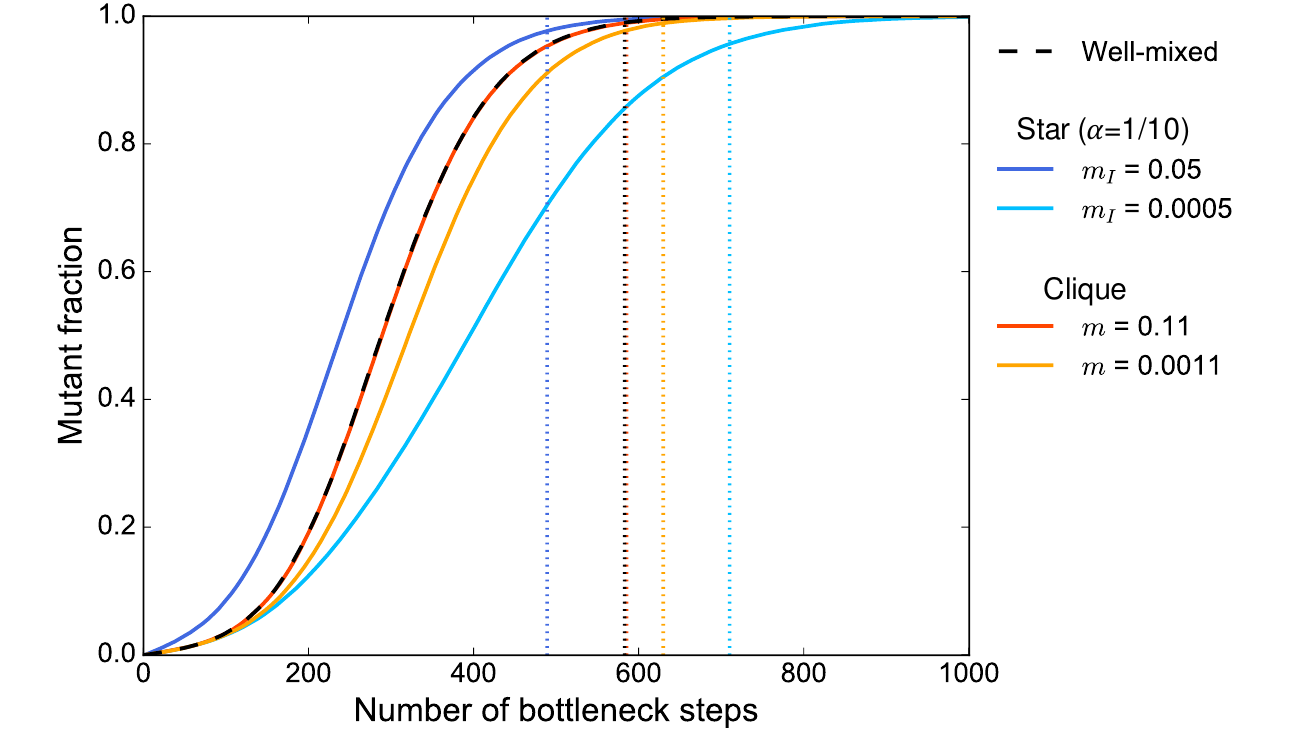}
\caption{\textbf{Mutant fraction growth leading to fixation}. The mutant fraction in the total population is shown versus the number of bottleneck steps for the well-mixed population, the star and the clique in different migration regimes. We consider a star with $D=5$ demes, and $K=1000$ individuals per deme on average at the bottleneck, for migration asymmetry $\alpha=1/10$, leading to strong suppression of selection (see Fig.~\ref{star_fprobas_etimes_BP}). We also consider a clique with the same number of demes of the same size. The well-mixed case is shown for comparison, with $KD=5000$ individuals at the bottleneck. In all cases, we start with one mutant of fitness $f_M=1+s$ placed uniformly at random at a bottleneck, all other individuals being wild-types with fitness $f_W=1$. The growth phase duration is $t=5$, and the effective fitness advantage is set to $st=0.2$. Curves represent simulation results, averaged over 2000 to 5000 fixation trajectories. Vertical dotted lines show the average fixation times for these structures. Note that those corresponding to the clique with $m=0.11$ and to the star are superposed.}
\label{fraction_evo}
\end{center}
\end{figure}

Fig.~\ref{fraction_evo} shows the growth of mutant fraction in trajectories that end in mutant fixation, starting from one mutant in different spatial structures. For the migration asymmetry chosen, which leads to strong suppression of selection (see Fig.~\ref{star_fprobas_etimes_BP}), we observe that mutant fraction grows faster in the star with frequent migrations than in the well-mixed population. This is associated to the suppression of selection and to the acceleration of extinction obtained in this case (see Fig.~\ref{star_fprobas_etimes_BP}). Meanwhile, the clique with frequent migrations behaves like the well-mixed population, corroborating their similarities shown for fixation probabilities and for extinction times (see Section~\ref{exttimecirc}). For less frequent migrations, the growth of mutant fraction is slower in structured populations (clique and star) than in the well-mixed population: in this regime, spatial structure slows down the dynamics. 

\begin{figure}[h!]
\begin{center}
\includegraphics[width=\textwidth]{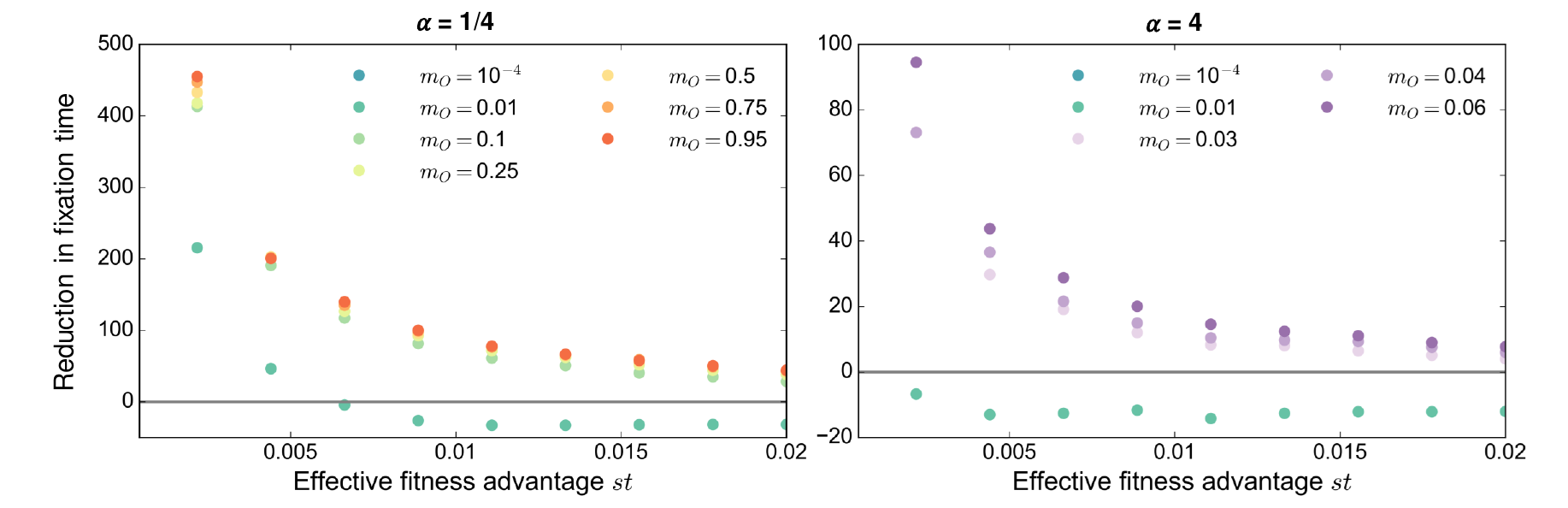}
\caption{\textbf{Reduction of the average mutant fixation time in the star.} The difference between the average fixation time (in numbers of dilution steps) in the well mixed population and that in the star is plotted versus the effective fitness advantage $st$ of the mutant. As in Fig.~\ref{star_fprobas_etimes_BP}, we consider a star with $D=5$ demes, and $K=1000$ individuals per deme on average at the bottleneck, for migration asymmetries $\alpha=1/4$ (left) and $\alpha=4$ (right).  We start with one mutant of fitness $f_M=1+s$ placed uniformly at random at a bottleneck, all other individuals being wild-types with fitness $f_W=1$. The growth phase duration is $t=5$. Markers represent simulation results averaged over 100 000 realizations. The well-mixed case, taken as reference, corresponds to a population with $KD=5000$ individuals at the bottleneck, initialized with one mutant. }
\label{star_ftimes}
\end{center}
\end{figure}

Fig.~\ref{star_ftimes} shows that, for frequent migrations, the average mutant fixation time is smaller in the star than in the well-mixed population. This further confirms the acceleration of the dynamics observed in this regime, in Fig.~\ref{star_fprobas_etimes_BP} for extinction times, and in Fig.~\ref{fraction_evo} for the growth of mutant fraction in trajectories yielding fixation.

\newpage

\subsection{When all demes contribute by $K$ on average}

The previous subsections focused on the case where all demes have the same average bottleneck size $K$, i.e. $\sum_j m_{ji}=1$ for all $i$. The branching process approximation can also be used in the case where all demes contribute by $K$ individuals on average to each bottleneck, but can have different average bottleneck sizes. In that case, $\sum_j m_{ij}=1$ for all $i$. The generating function's $i$-th component keeps the form of Eq.~\ref{generating_bin}, but migrating probabilities $m_{ij}$ are replaced by $\mu_{ij}=m_{ij}/\sum_k m_{ki}$, as in the previous section:
\begin{equation}
    f_i(\mathbf{x})=\prod_{j=1}^D \left( 1 - (1-x_j) \dfrac{\lambda}{N'_i}\mu_{ij}\right)^{N'_i} \,.
\end{equation}
All the calculations performed assuming $\sum_j m_{ji}=1$ for all $i$ can be adapted simply replacing $m_{ij}$ by $\mu_{ij}$, yielding predictions for fixation probabilities and average fixation time for different graph structures. 

\begin{figure}[htbp]
\begin{center}
\includegraphics[scale=0.45]{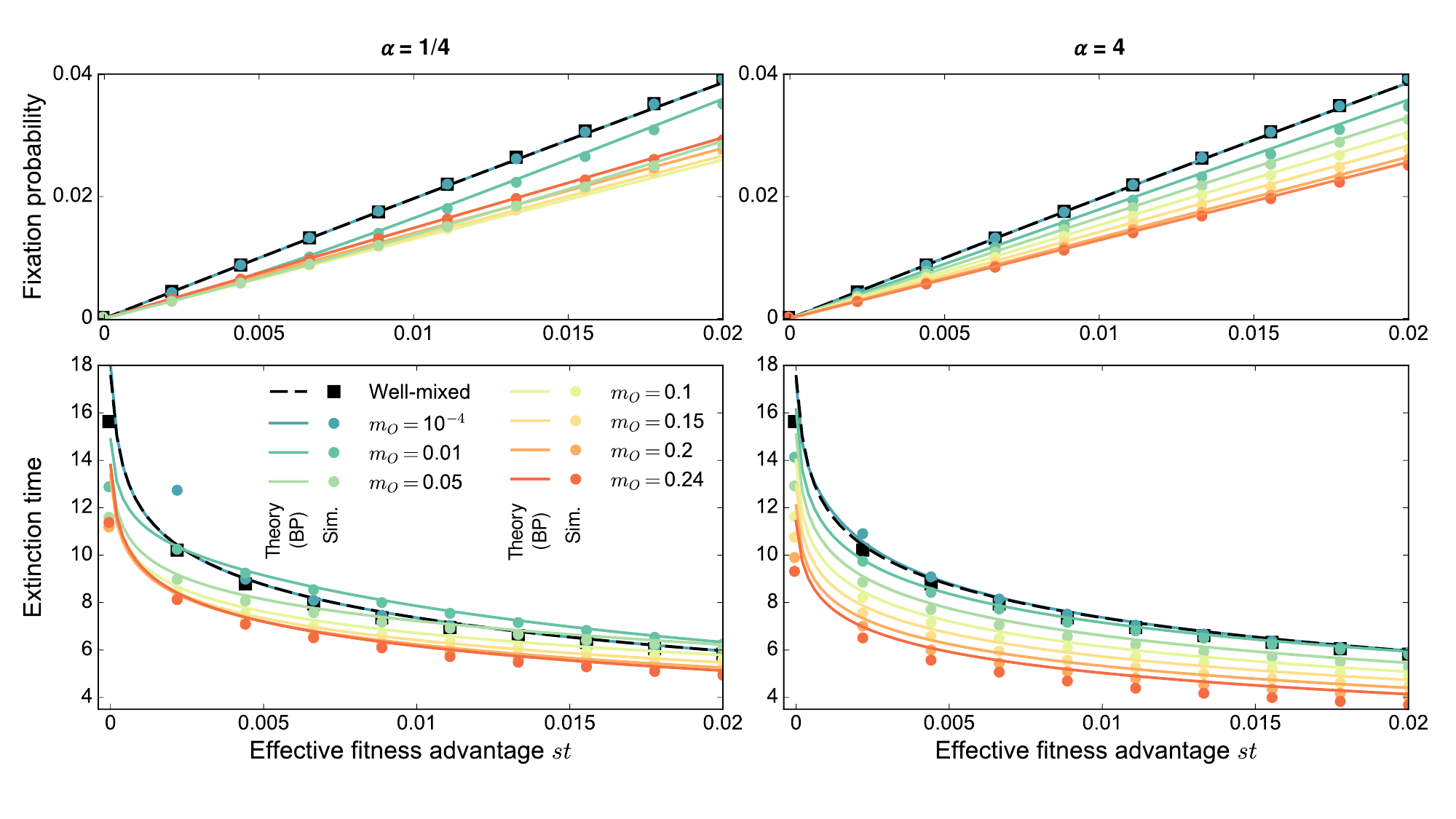}
\caption{\textbf{Mutant fixation in the star when all demes contribute by $K$ on average.} Mutant fixation probability (top) and average extinction time (bottom, in numbers of dilution steps) are plotted versus the effective fitness advantage $st$ of the mutant. We consider a star with $D=5$ demes, and all demes contribute on average by $K=1000$ individuals to each bottleneck. Migration asymmetries are $\alpha=1/4$ (left) and $\alpha=4$ (right). We start with one mutant of fitness $f_M=1+s$ placed uniformly at random at a bottleneck, all other individuals being wild-types with fitness $f_W=1$. The growth phase duration is $t=5$. Markers represent simulation results (``Sim.''), averaged on 1 million realizations. Lines are theoretical predictions from our branching process (``BP'') approach. The well-mixed case is shown for comparison, with simulations performed for a well-mixed population with $KD=5000$ individuals at the bottleneck, initialized with one mutant.  
The convention $\sum_j m_{ij} =1$ constrains the value of $m_O$ to be between 0 and 1/4.}
\label{star_fprobas_etimes_BP_eqcon}
\end{center}
\end{figure}

In Fig.~\ref{star_fprobas_etimes_BP_eqcon}, we show both analytical predictions and simulation results for the star graph with $D=5$ and equally contributing demes.  Note that corresponding results in the case where all demes start at the same bottleneck size on average are shown in Fig.~\ref{star_fprobas_etimes_BP}. 

\newpage
\section{Different models for serial dilution in structured populations}
\label{diff_models}

Several models can be considered to describe migration and dilution in a structured population. Here, we define three of them, which rely on distinct sampling schemes which involve different fluctuations in the number of individuals that compose each new bottleneck state. This could affect the evolutionary dynamics. Let us compare these models, in the convention used in the main text where demes have the same average size $K$ at the bottleneck, i.e. $\sum_j m_{ji} = 1$ for all $i$.

\paragraph{Binomial samplings for migration and dilution.} This model is the main one studied in this paper, and is detailed in the Methods in the main text. Starting from $D$ demes with mutant fractions $x_1,\dots,x_D$, we begin with a growth phase at the end of which the mutant ratios become $x_1',\dots,x_D'$ in demes of large sizes $N_1',\dots,N_D'$. Then, we proceed with the dilution and migration step by drawing the number of migrating mutants (resp. wild-types) from deme $i$ to $j$ in a binomial distribution with $N_i'$ trials and probability of success $K/N_i' x_i' m_{ij}$ (resp. $K/N_i' (1-x_i') m_{ij}$). This model is based on independent binomial samplings for each type, for any migration itinerary. The bottleneck size of each deme fluctuates around the average value $K$.

\paragraph{Multinomial samplings for migration and dilution.}
A variant uses multinomial samplings at the migration and dilution step, after the growth phase. For any deme $i$, the numbers of incoming individuals from all demes (including mutants and wild-types) are sampled from a multinomial distribution with $K$ trials and probabilities $(x_1' m_{1i}, (1-x_1') m_{1i}, \dots, x_D' m_{Di}, (1-x_D') m_{Di})$. Note that these probabilities sum to 1 since $\sum_j m_{ji}=1$. In this model, the deme sizes are exactly equal to $K$ at bottleneck. Multinomial samplings are performed independently  for each deme. Applied to a well-mixed population, this model reduces to the Wright-Fisher model.

\paragraph{Local binomial samplings for dilution, after migration.} In this model, migrations are performed as a deterministic step, before sampling, exactly as in a structured Wright-Fisher model~\cite{Burden18}. After the growth phase, the mutant ratios are $x_1',\dots,x_D'$. After migration (and before dilution), the local mutant ratios become $\tilde{x}_1,\dots,\tilde{x}_D$ where $\tilde{x}_i = \sum_j m_{ji}x_j'$ for all $i$. Note that this migration step does not take into account the actual sizes of the demes, essentially assuming that they are all infinite. Then, dilution is performed locally in each deme by sampling the number of mutant individuals present in deme $i$ at the next bottleneck from a binomial distribution with $K$ trials and probability of success $\tilde{x}_i$. The number of wild-types in deme $i$ at the next bottleneck is then taken so that deme $i$ has exactly size $K$. Applied to a well-mixed population, this model also reduces to the Wright-Fisher model.

\begin{figure}[htb]
\begin{center}
\includegraphics[scale=0.45]{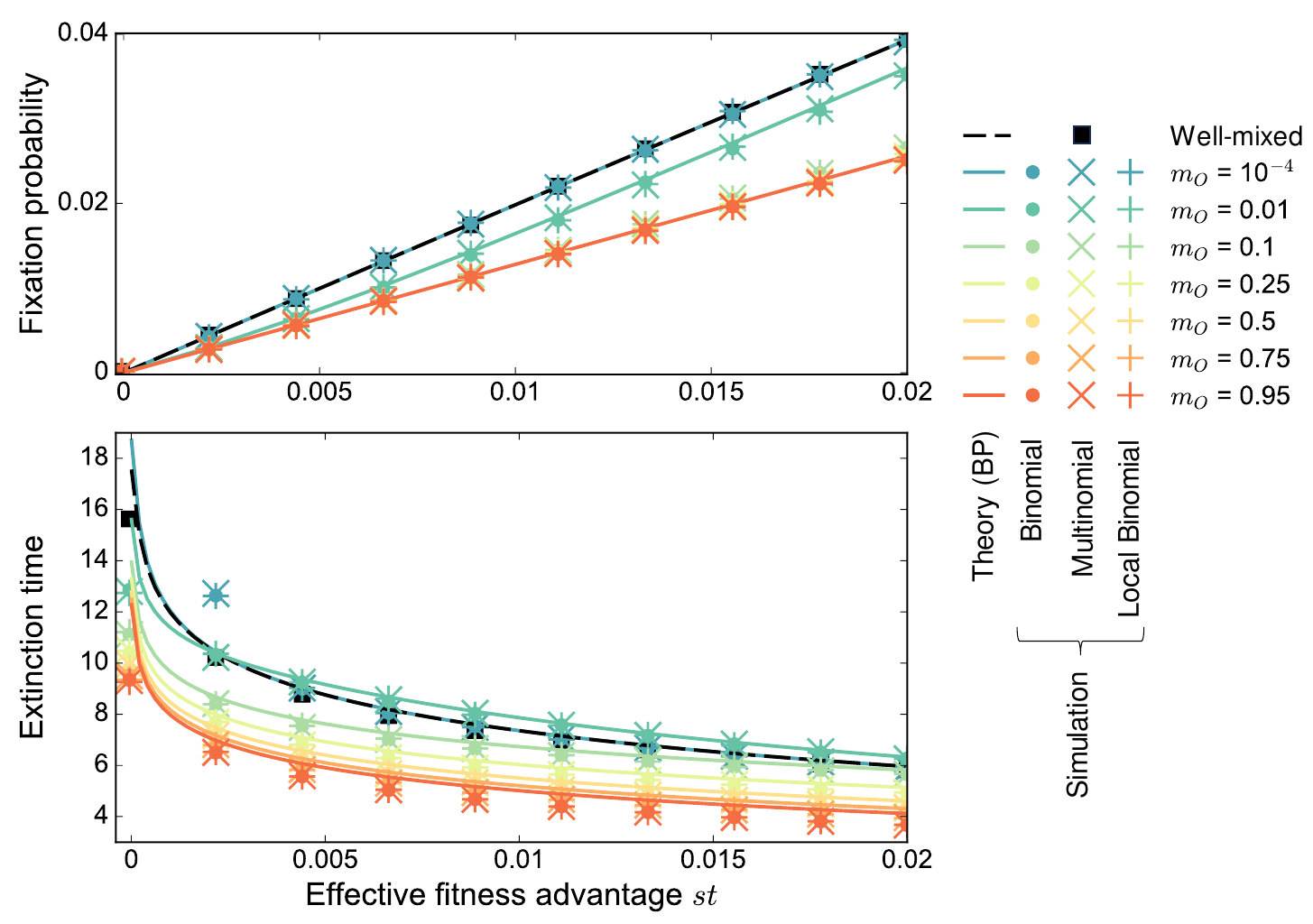}
\caption{\textbf{Mutant fixation in the star for different models of serial dilution.} Mutant fixation probability (up) and average extinction time (bottom, in numbers of dilution steps) are plotted versus the effective fitness advantage $st$ of the mutant. We consider a star with $D=5$ demes, and $K=1000$ individuals per deme at the bottleneck (either exactly or on average depending on the model), for migration asymmetry $\alpha=1/4$. The growth phase duration is $t=5$.  We start with one mutant of fitness $f_M=1+s$ placed uniformly at random at a bottleneck, all other individuals being wild-types with fitness $f_W=1$. Markers are simulation results (``Sim.'') obtained from 1 million realizations. The three different markers correspond to three different models defined in Section~\ref{diff_models}, that use binomial samplings for each migration itinerary and each type (``Binomial'') , multinomial samplings for each deme (``Multinomial''), or binomial samplings for each deme after a deterministic migration step (``Local Binomial'') . Lines are theoretical predictions obtained through a branching process (``BP'') approach. The well-mixed case is shown for comparison, with simulations performed for a well-mixed population with $KD=5000$ individuals at the bottleneck, initialized with one mutant. }
\label{star_fprobas_BP_allsamplings}
\end{center}
\end{figure}

\paragraph{Comparison.} The branching process approach can be applied to these three models. Under the assumption $st \gg 1/K$, we find that the first three order terms in the perturbative expansion in $st$ of the generating function, detailed for the first model in Eq.~\ref{generating_bin}, are the same for all models. Therefore, all those models should yield the same results in the domain of validity of the branching process analysis. This is indeed confirmed by simulations, as shown in Fig.~\ref{star_fprobas_BP_allsamplings}.

\newpage

\section{Dirichlet cliques}
\label{Dirichlet_clique}

To go beyond graphs with high symmetry like the clique and the star, we consider graphs where migration probabilities are drawn in Dirichlet distributions. Indeed, using these probability distributions ensure that our migration probabilities satisfy $\sum_i m_{ij}=1$ for all $i$, and allow us to tune migration asymmetry. Here, we give the definition of the Dirichlet distribution, as well as some examples, and we explain how we use it to generate migration probabilities.

\subsection{Definition of the Dirichlet distribution}

The Dirichlet distribution of order $D\geq 2$ with parameters $\eta_1,...,\eta_D$ has the following probability density function for $\mathbf{x}=(x_1,...,x_D) \in [0,1]^D $ in the $D-1$ simplex (such that $\sum_i x_i = 1$):
\begin{equation}
    \Phi_{(\eta_1,...,\eta_D)}(\mathbf{x})=\dfrac{1}{B(\eta_1,...,\eta_D)} \prod_{i=1}^D x_i^{\eta_i - 1}\,,
\end{equation}
where $B$ is the Euler beta function, which can be expressed in terms of the gamma function $\Gamma$ as
\begin{equation}
    B(\eta_1,...,\eta_D)=\dfrac{\prod_{i=1}^D \Gamma (\eta_i)}{\Gamma \left(\sum_{i=1}^D \eta_i \right)}\,.
\end{equation}

The parameters $\eta_1,...,\eta_D$ impact both the mean values of the sampled variables and their variances and covariances. Indeed, random variables $X_1,...,X_D$ sampled from the Dirichlet distribution have the following means and covariances:
\begin{align}
    \mathbb{E}[X_i] &= \tilde{\eta}_i\\
    \textrm{Var}[X_i]&=\dfrac{\tilde{\eta}_i(1-\tilde{\eta}_i)}{\sum_j \eta_j+1} \\
    \textrm{Cov}[X_i,X_j]&= \dfrac{- \tilde{\eta}_i \tilde{\eta}_j}{\sum_j \eta_j+1} \quad \text{ for } i \neq j \,,
\end{align}
where $\tilde{\eta}_i=\eta_i/\sum_j \eta_j$. Thus, on average, the higher $\eta_i$ is with respect to the other coefficients $\eta_j$ with $j\neq i$, the closer $x_i$ is to one. Moreover, as the sum $\sum_j \eta_j$ of the coefficients $\eta_j$ grows, the variances and covariances of the variables $X_1,...,X_D$ decrease. 

\subsection{Examples of Dirichlet distributions}

\begin{figure}[htb!]
\begin{center}
\includegraphics[scale=0.45]{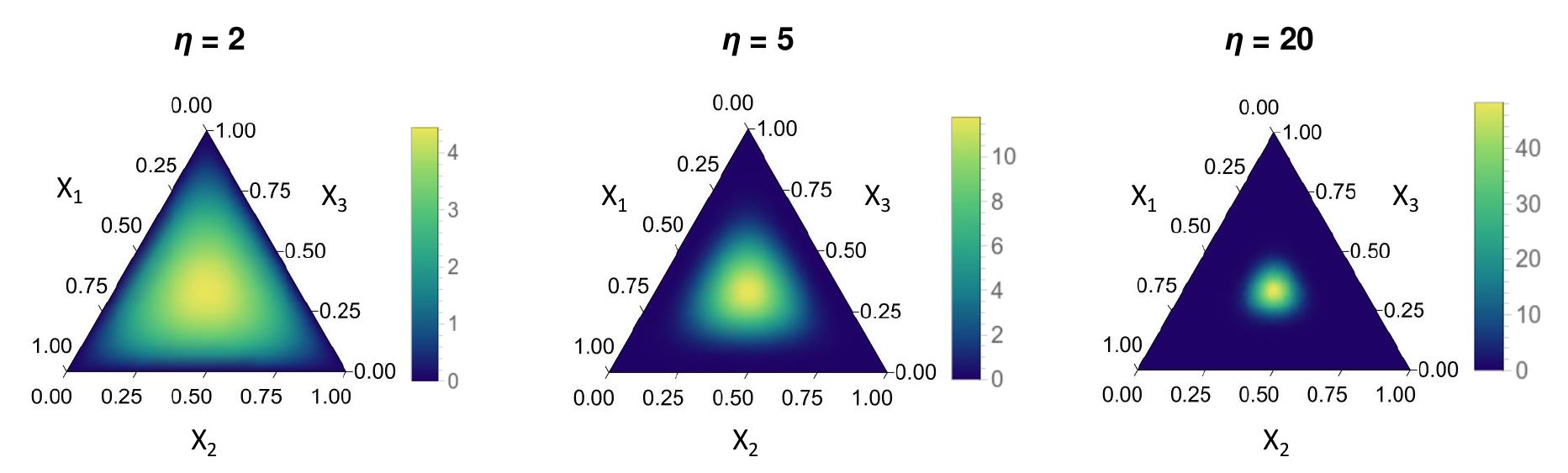}
\caption{Probability density function of random variables $(X_1,X_2,X_3)$ sampled from a Dirichlet distribution with parameters all equal to $\eta$. The probability density function is represented on a ternary plot, where point coordinates are obtained by projecting along lines parallel to the triangle edges. The mean values of $X_1, X_2, X_3$ are always equal to $1/3$, and the density concentrates around the central point as $\eta$ grows.}
\label{Dirichlet_eta}
\end{center}
\end{figure}

\paragraph{When all parameters are equal.}
If all parameters $\eta_1,...,\eta_D$ are equal to $\eta$, then $X_1,...,X_D$ all have the same mean $1/D$. When $\eta=1$, the distribution is uniform on the simplex. If $\eta$ grows above 1, the variances and covariances of $X_1,...,X_D$ decrease, and their probability density concentrates around the mean. Conversely, as $\eta$ becomes smaller, variances grow and the probability density concentrates around small and large values of $x_1,...,x_D$, near the simplex boundary. Fig.~\ref{Dirichlet_eta} illustrates the impact of increasing $\eta$ for $D=3$, where the random variables $X_1, X_2, X_3$ have average value $1/3$. 

\begin{figure}[htb!]
\begin{center}
\includegraphics[scale=0.45]{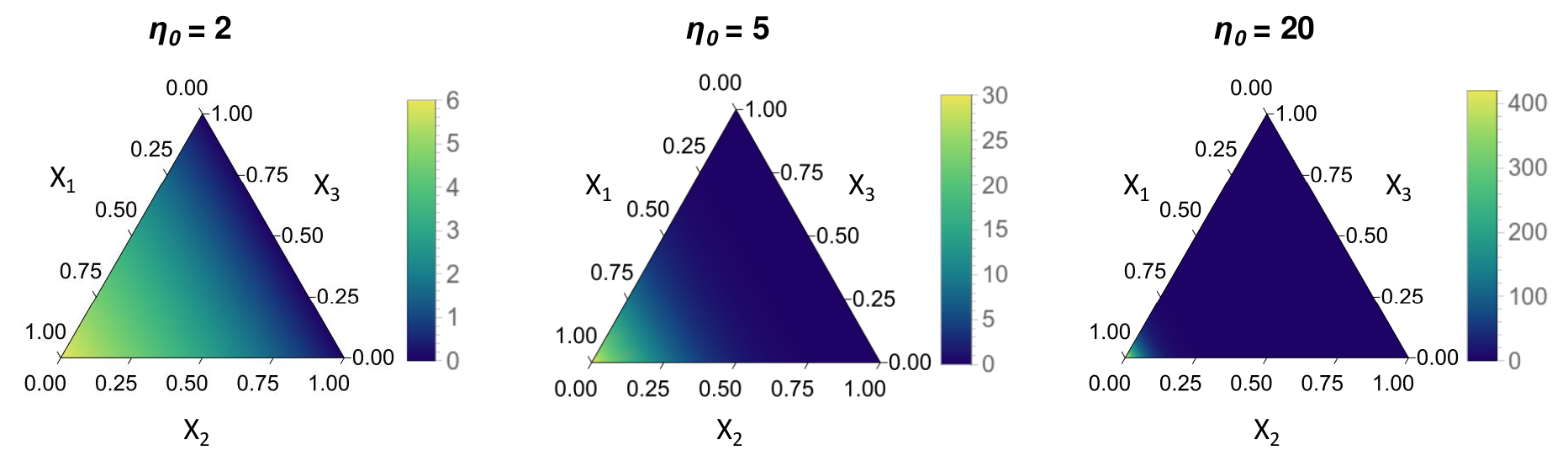}
\caption{Ternary plot of the probability density function of random variables $(X_1,X_2,X_3)$ sampled from a Dirichlet distribution with parameters $(\eta_0, 1, 1)$. As $\eta_0$ grows, the probability density concentrates around the left corner, which corresponds to $(x_1,x_2,x_3)=(1,0,0)$.}
\label{Dirichlet_eta0}
\end{center}
\end{figure}

\paragraph{When one parameter is larger than others.}
Consider a Dirichlet distribution of parameters $\eta_1\equiv\eta_0,\,\eta_2=1,\,...,\,\eta_D=1$ with $\eta_0>1$. The variables $X_1,...,X_D$ have mean values
\begin{align}
    \mathbb{E}[X_1]&=\dfrac{\eta_0}{\eta_0 + D-1}\,, \\
    \mathbb{E}[X_j]&=\dfrac{1}{\eta_0 + D-1} \quad \text{ for }j \geq 2\,.
\end{align}
As $\eta_0$ becomes larger, the mean of $X_1$ grows towards 1, while the mean of $X_j$ for $j>1$ decreases towards 0. If $\eta_0>D-1$, when $\eta_0$ grows, the variances of $X_1,...,X_D$ decrease towards 0, while their covariances increase towards 0. Therefore, the probability density concentrates around the mean values. An example is shown for $D=3$ in Fig.~\ref{Dirichlet_eta0}.

\newpage
\subsection{Application: Dirichlet cliques}

In Fig.~\ref{clique_random}, for $D=5$, we generated graphs with migration probabilities drawn in Dirichlet distributions, which we call Dirichlet cliques. For any deme $j$, we sampled the incoming migration probabilities $(m_{1j},m_{2j},m_{3j},m_{4j},m_{5j})$ from a Dirichlet distribution. 

If all demes are equivalent (top panel of Fig.~\ref{clique_random}), we take all Dirichlet parameters equal to $\eta$. All migration probabilites then have mean $1/5$. Sampling a Dirichlet clique with $\eta$ very large yields a Dirichlet clique very close to the standard clique, with $m=1/5$.

If deme $1$ is advantaged (bottom panel of Fig.~\ref{clique_random}), then for any deme $j$, the incoming migration probabilities $(m_{1j},m_{2j},m_{3j},m_{4j},m_{5j})$ are sampled from a Dirichlet distribution with parameters $(\eta_0, 1, 1, 1, 1)$. This means that all demes are more likely to receive migrants from deme $1$ than from the others.

\bibliographystyle{unsrt}
\bibliography{biblio}

\end{document}